\documentclass[journal,rotate]{IEEEtran}

\usepackage{graphicx}  

\usepackage{subfigure} 

\usepackage{amsmath}   

\usepackage{amssymb}

\usepackage{algorithmic}
\usepackage{algorithm}
\usepackage{rotating}

\newtheorem{myfunc}{Function}

\begin{document}
%
\title{Projection-Based and Look Ahead Strategies \\ for Atom Selection}

%
%

\author{Saikat Chatterjee, Dennis Sundman, Mikko Vehkaper\"{a}, Mikael Skoglund \thanks{The authors are with Communication Theory Laboratory, School of Electrical Engineering, KTH - Royal Institute of Technology, Sweden. 
Emails: $\{$sach, denniss, mikkov, skoglund$\}$@kth.se}}
\maketitle

\begin{abstract}
In this paper, we improve iterative greedy search algorithms in which atoms are selected serially over iterations, i.e., one-by-one over iterations. For serial atom selection, we devise two new schemes to select an atom from a set of potential atoms in each iteration. The two new schemes lead to two new algorithms. For both the algorithms, in each iteration, the set of potential atoms is found using a standard matched filter. In case of the first scheme, we propose an orthogonal projection strategy that selects an atom from the set of potential atoms. Then, for the second scheme, we propose a look ahead strategy such that the selection of an atom in the current iteration has an effect on the future iterations. The use of look ahead strategy requires a higher computational resource. To achieve a trade-off between performance and complexity, we use the two new schemes in cascade and develop a third new algorithm. Through experimental evaluations, we compare the proposed algorithms with existing greedy search and convex relaxation algorithms.             
\end{abstract}


%
\IEEEpeerreviewmaketitle

\section{Introduction}

A linear underdetermined system model based sparse signal estimation problem has attracted much attention in the current literature. The sparse signal estimation problem 
has many applications, for example, in linear regression \cite{Miller_Subset_Selection_in_Regression}, communication \cite{Fletcher_OnOff_ISIT}-\cite{Gastpar_Sastry_Distributed_sensor_perception}, multimedia \cite{Daudet_Molecular_Matching_Pursuit}-\cite{Sastry_Face recognition}, and recently in compressive sampling (CS) \cite{Donoho_Compressed_sensing}-\cite{CS_introduction_Candes_Wakin}.
The algorithms proposed for solving such a problem may be categorized into three broad classes: convex relaxation, \cite{Donoho_Compressed_sensing}-\cite{Candes_Romberg_L1_Magic_Toolbox}, Bayesian inference \cite{Ji_Xue_Carin_Bayesian_CS}-\cite{Wipf_Rao_Sparse_Bayesian_learning}, and iterative greedy search \cite{Mallat_Matching_pursuits}-\cite{Blumensath_Davies_Iterative thresholding_2}.
The iterative greedy search (IGS) algorithms are of rising interest for solving large dimensional sparse signal estimation problems due to their algorithmic simplicity and lower complexity.
 
Generally an IGS algorithm constructs a support set of the underlying sparse signal vector through iterations. 
In the literature, the support set is referred to as the set of indices corresponding to coordinates of non-zero elements of the sparse signal vector. Also the columns of the measurement matrix (or dictionary matrix) are referred to as atoms. Naturally, the support set's elements are the indices of atoms associated with the non-zero elements of the sparse vector. In the underdetermined setup, an atom-selection process for constructing the support set is a critical task.  
Depending on atom selection approaches, the IGS algorithms may be categorized into two classes: serial (or sequential) and parallel (or simultaneous).
Prominent examples of the serial atom selection based IGS algorithms are orthogonal matching pursuit (OMP) \cite{Mallat_Orthogonal_Matching_pursuits}-\cite{Pati_Orthogonal_Matching_pursuits} and orthogonal least squares (OLS) \cite{Chen_Orthogonal_least_squares}-\cite{Natarajan_SIAM_Paper}. For both the algorithms, a single atom is selected from a set of potential atoms in each iteration and then the atom's index is inducted into the support set under progressive construction. Therefore, the cardinality of the support set under construction is increased one-by-one through iterations serially\footnote{Whilst both the OMP and OLS algorithms are structurally similar, there exists a subtle difference between them in the way an atom is selected in each iteration. A clear exposition on the difference between these two algorithms can be found in \cite{Blumensath_Difference_on_OMP_and_OLS}. The reader is encouraged to see Appendix~\ref{subsec:Serial_Atom_Selection_Based_OMP_and_OLS}.}. 
At the end of all iterations, the algorithms provide a support set of full cardinality.
On the other hand, for the parallel atom selection based IGS algorithms, such as subspace pursuit (SP) \cite{Dai_Subspace_pursuit} and CoSaMP \cite{Needell_Tropp_CoSaMP}, a support set with a fixed (pre-determined) cardinality is estimated in each iteration\footnote{We briefly describe serial atom selection based OMP and OLS algorithms in Appendix~\ref{subsec:Serial_Atom_Selection_Based_OMP_and_OLS} and parallel atom selection based SP in Appendix~\ref{subsec:Parallel_Atom_Selection_Based_SP}.}. A fixed number of atoms is selected simultaneously in each iteration to form the support set of full cardinality and the estimate of the underlying support set is refined through iterations. 
In practice, IGS algorithms provide different support set estimation performances (in turn different sparse signal estimation performances). For selecting (or detecting) an atom in a computationally efficient manner, the use of a matched filter remains prevalent in existing IGS algorithms.

In this paper, we develop new IGS algorithms to provide a better estimate of the underlying support set at the expense of computational complexity. Our interest is to deal with a highly underdetermined system that may arise in many applications, for example in a highly under-sampled CS. For such a underdetermined system, the reliability of the matched filter degrades. At the expense of complexity, we endeavor for improving atom selection performance beyond the scope of a matched filter. In the new algorithms, a trade-off between complexity and performance can be achieved by adjusting user defined tunable parameters.  

We develop the new algorithms using the architectures of OMP and OLS algorithms where atom selection is performed serially over iterations. For the new algorithms, in each iteration, a set of potential atoms is chosen and then a single atom is finally selected from the set of potential atoms. We propose new atom selection strategies and incorporate them in the OMP and OLS architectures. Strategically, the new algorithms use a two stage mechanism: (a) first, the standard matched filter is used to choose a set of potential atoms from all the competing atoms, and (b) second, an atom is finally selected from the set of potential atoms by using the new selection strategies. In the first stage, the cardinality of the set of potential atoms is pre-determined. The use of matched filter in the first stage helps to reduce search space for the new selection strategies invoked in the second stage. 

For the standard OMP, in each iteration, the matched filter is used to select an atom from all the competing atoms. In a highly underdetermined setup, atoms are more correlated and hence the atom selection performance using matched filter degrades. This performance degradation is due to fact that an underlying atom is subjected to a considerable interference from other closely correlated atoms. 
To improve atom selection, we propose to use the same matched filter for choosing a set of potential atoms and then select an atom from the set using an \emph{orthogonal projection} strategy. The new algorithm is referred to as the projection based OMP (POMP) where the use of orthogonal projection is motivated by its existing use in parallel atom selection based IGS algorithms (such as in SP and CoSaMP). 
The orthogonal projection invokes signal estimation in least-squares (LS) sense and hence its use adds more strength on the top of matched filter.

Next, for the standard OLS, an atom is selected in each iteration such that the selection leads to the minimum norm of a fitting residual (LS residual) for that iteration. 
The selection of an atom in the current iteration does not depend on the final performance of the OLS algorithm when all the iterations would have been finished. Therefore, the selection of a new atom in the current iteration is blind to its effect on the future iterations in the sense of minimizing the final fitting residual norm. To overcome the shortcoming of blindness in OLS, we propose to use a \emph{look ahead} selection strategy \cite{Chatterjee_Sundman_Skoglund_4}. In the current iteration, an atom is selected from the set of potential atoms by evaluating its effect on the final performance measure in the sense of minimizing the norm of fitting residual at the end of all future iterations. The new algorithm is referred to as look ahead OLS (LAOLS). Use of a look ahead strategy is shown to provide a better performance, but at the expense of a higher computation.

Further, we combine the orthogonal projection-based and look ahead atom selection strategies in cascade for developing an IGS algorithm which provides a trade-off between computational complexity and reconstruction performance.  

Considering compressive sampling (CS) as the application, we experimentally evaluate the new algorithms and compare their performances vis-a-vis several existing algorithms, such as OMP, OLS, SP and convex relaxation algorithms. For experiments, we endeavor for CS reconstruction of Gaussian and binary sparse signals at varying under-sampling rates and measurement noise levels.  
  
Notations: Let $\mathbf{A} \in \mathbb{R}^{M \times N}$, $\mathbf{x} \in \mathbb{R}^{N}$, and $\mathcal{I} \subset \{1,2,\ldots,N\}$. We denote $|\mathcal{I}|$ and $\overline{\mathcal{I}}$ as the cardinality and complement of the set $\mathcal{I}$, respectively. The matrix $\mathbf{A}_{\mathcal{I}} \in \mathbb{R}^{M \times |\mathcal{I}|}$ consists of the columns of $\mathbf{A}$ indexed by $i \in \mathcal{I}$, and $\mathbf{x}_{\mathcal{I}}\in \mathbb{R}^{|\mathcal{I}|}$ is composed of the components of $\mathbf{x}$ indexed by $i \in \mathcal{I}$.  Also we denote $(.)^{t}$ and $(.)^{\dag}$ as transpose and pseudo-inverse, respectively.

\subsection{Preliminaries of Compressive Sampling}
\label{subsec:CS_and_Orthogonal_Greedy_Pursuits}

Let us state the standard CS problem where we acquire a $K$-sparse signal $\mathbf{x} \in \mathbb{R}^{N}$ via the linear measurements
\begin{eqnarray}
\mathbf{y} = \mathbf{A} \mathbf{x} + \mathbf{w},
\label{eq:CS_Standard_Problem}
\end{eqnarray}
where $\mathbf{A} \in \mathbb{R}^{M \times N}$ is a matrix representing the sampling system, $\mathbf{y} \in \mathbb{R}^{M}$ represents a vector of measurements and $\mathbf{w} \in \mathbb{R}^{M}$ is additive noise representing measurement errors. A $K$-sparse signal vector consists of at most $K$ non-zero scalar components. For the setup $K<M<N$ (underdetermined system of linear equations), the task is to reconstruct $\mathbf{x}$ from $\mathbf{y}$ as $\hat{\mathbf{x}}$. With no a-priori statistical knowledge of $\mathbf{x}$ and $\mathbf{w}$, the objective in CS is to strive for a reduced number of measurements ($M$) as well as achieving a good reconstruction quality. Note that, in practice, we may wish to acquire a signal $\mathbf{x}$ that is sparse in a known orthonormal basis and the concerned problem can be recast as (\ref{eq:CS_Standard_Problem}).

For a sparse signal vector $\mathbf{x}=\left[ x_{1},\,x_{2},\ldots,x_{N}\right]^{t}$, the support set $\mathcal{I} \subset \{1,2,\ldots,N\}$ is defined as $\mathcal{I} = \{ i \, : \, x_{i} \neq 0 \}$. For a $K$-sparse vector $\mathbf{x} \in \mathbb{R}^{N}$, $|\mathcal{I}|=\|\mathbf{x}\|_{0} \leq K$. In this paper, we assume that $|\mathcal{I}|=K$. Denoting the $i$'th column (atom) of the measurement matrix $\mathbf{A}$ as $\mathbf{a}_{i}$, note from~(\ref{eq:CS_Standard_Problem}) that
\begin{eqnarray}
\mathbf{y} = \sum_{i \in \mathcal{I} } x_{i} \, \mathbf{a}_{i} + \mathbf{w} = \mathbf{A}_{\mathcal{I}} \, \mathbf{x}_{\mathcal{I}} + \mathbf{w},
\end{eqnarray}
where $\mathbf{x}_{\mathcal{I}} \in \mathbb{R}^{K}$ and $\mathbf{A}_{\mathcal{I}} \in \mathbb{R}^{M \times K}$.
From $\mathbf{y}$, if the underlying support set $\mathcal{I}$ can be estimated, then we can estimate the non-zero values of $\mathbf{x}$ (i.e. $\mathbf{x}_{\mathcal{I}}$) using a standard least squares (LS) solution (as $K<M$, we can use the pseudo-inverse). Therefore, a better estimate of the support set leads to a better reconstruction performance. However, if the matrix $\mathbf{A}_{\mathcal{I}}$ is not full column rank (or ill-conditioned) then the LS solution will be erroneous. This may happen in case of the matrices where the correlation between the columns are considerably high, for example, in a highly under-sampled CS where $M \ll N$. Recently, theoretical properties of such matrices are investigated in \cite{Mixon_Bajwa_Calderbank_ISIT2011}, \cite{Bajwaa_Calderbanka_Mixon_Journal}. 

Next we consider the issue of constructing the sensing matrix $\mathbf{A} \in \mathbb{R}^{M \times N}$. While it is possible to obtain deterministic constructions of $\mathbf{A}=\left\{ a_{i,j} \right\}$ holding a specific structure, at present the most efficient designs (i.e., those requiring minimum number of rows) rely on random matrix constructions where the $a_{i,j}$'s are assumed realizations of independent and identically distributed (i.i.d.) random variables. A standard method is to draw $a_{i,j}$'s independently from a Gaussian source (i.e., $a_{i,j} \sim \mathcal{N} \left(0, \frac{1}{M} \right)$) and then to scale the columns of $\mathbf{A}$ to unit-norm. Note that once $\mathbf{A}$ is constructed, it remains fixed and made known to a CS reconstruction algorithm. In the following sections, we develop the new IGS algorithms.

\section{Projection Based Orthogonal Matching Pursuit}
\label{sec:POMP}

Using the algorithmic structure of OMP, we develop projection based OMP (POMP) where the selection of an atom in the current iteration is carried out using an orthogonal projection method. 

For developing the POMP algorithm, we first need to describe the orthogonal projection based selection method in Algorithm~\ref{alg:Proj_atom_selection}.
\begin{algorithm}[ht!]
\caption{: Projection-Based Atom-index Selection}\label{alg:Proj_atom_selection}
\mbox{Input: }
\begin{algorithmic}[1]
\STATE $\mathbf{A}$, $\mathbf{y}$;
\STATE Previous support set: $\mathcal{I}$;
\STATE Set of potential atoms' indices: $\mathcal{I}_{(p)}$; 
\end{algorithmic}
\mbox{Assumption: }
\begin{algorithmic}[1]
\STATE $\mathcal{I}_{(p)} \cap \mathcal{I} = \emptyset$, $\hat{\mathbf{x}} \in \mathbb{R}^{N}$; 
\end{algorithmic}
\mbox{Execution: }
\begin{algorithmic}[1]
\STATE $\mathcal{I}_{(u)} \leftarrow \mathcal{I} \cup \mathcal{I}_{(p)}$; 
\STATE $\hat{\mathbf{x}}_{\mathcal{I}_{(u)}} \leftarrow \mathbf{A}_{\mathcal{I}_{(u)}}^{\dag} \mathbf{y}$; \hfill (Orthogonal projection)
\STATE $\hat{\mathbf{x}}_{\overline{\mathcal{I}}_{(p)}} \leftarrow \mathbf{0}$; \hfill (Zeroing the coordinates indexed by $\overline{\mathcal{I}}_{(p)}$)
\STATE $i \leftarrow \mathrm{index \,\, of \,\, the \,\, highest \,\, amplitude \,\, of \,\, \hat{\mathbf{x}}}$
\end{algorithmic}
\mbox{Output:}
\begin{algorithmic}[1]
\STATE The index $i$.  \hfill (Note: $i \in \mathcal{I}_{(p)}$)
\end{algorithmic}
\end{algorithm}
In Algorithm~\ref{alg:Proj_atom_selection}, the inputs are $\mathbf{A}$, $\mathbf{y}$, previous support set $\mathcal{I}$ and the set of potential atoms' indices $\mathcal{I}_{(p)}$. We refer to the previous support set $\mathcal{I}$ as the constructed support set in the previous iteration. The $\mathcal{I}_{(p)}$ is the set of potential atoms chosen using the standard matched filter in the current iteration. In Algorithm~\ref{alg:Proj_atom_selection}, a dummy set is formed as $\mathcal{I}_{(u)} = \mathcal{I} \cup \mathcal{I}_{(p)}$. Then the measurement vector $\mathbf{y}$ is projected on $\mathbf{A}_{\mathcal{I}_{(u)}}$ followed by constructing an estimate of $\hat{\mathbf{x}}$ where $\hat{\mathbf{x}}_{\overline{\mathcal{I}}_{(p)}} = \mathbf{0}$. The index of the most promising atom is then found using the standard criterion of finding the coordinate associated with the highest amplitude of $\hat{\mathbf{x}}$.  
Such an orthogonal projection based atom selection mechanism is used in parallel IGS algorithms, for example in SP \cite{Dai_Subspace_pursuit} and CoSaMP \cite{Needell_Tropp_CoSaMP}, albeit differently. To compare, the SP algorithm is described briefly in Appendix~\ref{subsec:Parallel_Atom_Selection_Based_SP}.

Using Algorithm~\ref{alg:Proj_atom_selection}, let us define the following function.
\begin{myfunc}
\label{func:Projection_based_atom-index_selection}
(Projection-based atom-index selection) Let $\mathbf{y} \in \mathbb{R}^{M}$, $\mathbf{A} \in \mathbb{R}^{M \times N}$, previous support set $\mathcal{I}$ and the set of potential atoms' indices $\mathcal{I}_{(p)}$ are given. Suppose $i$ denotes the index of the new atom selected in the current iteration. Then $i$ is the output of the following algorithmic function 
\begin{eqnarray}
i = \mathrm{proj \_ atom \_index}\left( \mathbf{A}, \mathbf{y}, \mathcal{I}, \mathcal{I}_{(p)} \right) 
\end{eqnarray}
where the above function executes the Algorithm~\ref{alg:Proj_atom_selection}.
\end{myfunc}

Using Function~\ref{func:Projection_based_atom-index_selection}, the main steps of the POMP algorithm are summarized in Algorithm~\ref{alg:POMP} where the support set is constructed serially.
\begin{algorithm}[ht!]
\caption{: POMP for CS Recovery}\label{alg:POMP}
\mbox{Input: }
\begin{algorithmic}[1]
\STATE $\mathbf{A}$, $\mathbf{y}$, $K$;
\STATE $L$ \hfill ($L$ potential atoms and $1 \leq L  \leq K$);
\end{algorithmic}
\mbox{Initialization: }
\begin{algorithmic}[1]
\STATE Iteration counter $k\leftarrow0$;
\STATE $\mathbf{r}_{0}\leftarrow\mathbf{y}$, $\mathcal{I}_{0}\leftarrow\emptyset$;
\end{algorithmic}
\mbox{Iterations: }
\begin{algorithmic}[1]
\REPEAT 
\STATE $k \leftarrow k+1$;
\STATE $\mathcal{I}_{(p)} \leftarrow$ $\{$indices of $L$ highest amplitudes of $\mathbf{A}^{t} \mathbf{r}_{k-1}\}$; \newline
(Assumption: chosen $L$ indices $\notin \mathcal{I}_{k-1}$) \label{step:Set_of_Potential_Atoms_POMP}
\STATE $i_{k} \leftarrow $ $\mathrm{proj \_ atom \_ index}$ $\left(\mathbf{A}, \mathbf{y}, \mathcal{I}_{k-1}, \mathcal{I}_{(p)} \right)$; \label{step:Proj_atom_index_POMP}
\STATE $\mathcal{I}_{k} \leftarrow \mathcal{I}_{k-1} \cup i_{k}$; \hfill (Note: $|\mathcal{I}_{k}|=k$) \label{step:Intermediate_Support_POMP}
\STATE $\mathbf{r}_{k} \leftarrow \mathbf{y}-\mathbf{A}_{\mathcal{I}_{k}} \mathbf{A}_{\mathcal{I}_{k}}^{\dag} \mathbf{y}$; \hfill (Orthogonal projection)	\label{step:Proj_Residue_POMP}
\UNTIL $( (\| \mathbf{r}_{k} \|_{2} > \| \mathbf{r}_{k-1} \|_{2}) \,\, \mathrm{or}\,\, (k>K) )$
\STATE $k \leftarrow k-1$; \hfill (Previous iteration)
\end{algorithmic}
\mbox{Output:}
\begin{algorithmic}[1]
\STATE $\hat{\mathbf{x}} \in \mathbb{R}^{N}$, satisfying $\hat{\mathbf{x}}_{\mathcal{I}_{k}} = \mathbf{A}_{\mathcal{I}_{k}}^{\dag} \mathbf{y}$ and $\hat{\mathbf{x}}_{\overline{\mathcal{I}}_{k}} = \mathbf{0}$.
\end{algorithmic}
\end{algorithm}
In the POMP algorithm, $L$ is a user defined positive integer parameter that decides how many atoms need to be checked as potential atoms in each iteration.
The algorithm starts with an initial empty support set $\mathcal{I}_{0}=\emptyset$ and an initial residual $\mathbf{r}_{0}=\mathbf{y}$. At the $k$'th iteration stage, it uses the standard matched filter to choose the set of $L$ potential atoms' indices (in step~\ref{step:Set_of_Potential_Atoms_POMP}), orthogonal projection method to select a new atom's index from the set (in step~\ref{step:Proj_atom_index_POMP}), inducts the selected index into the intermediate support set under construction (in step~\ref{step:Intermediate_Support_POMP}), solves a LS problem with the intermediate support set and produces a new residual by subtracting the LS fit (in step \ref{step:Proj_Residue_POMP}). Given the sparsity level $K$, the algorithm usually executes $K$ iterations and forms a support set of cardinality $K$ at the end. Note that the support set cardinality is increased one-by-one by selecting a new atom's index in each iteration. We also have used a stopping criterion that ensures a decreasing trend of residual norm over iterations and the maximum allowable cardinality of the constructed support set. We assume that the sparsity level $K$ is known a-priori like the cases of OMP (see Algorithm 3 of \cite{Tropp_OMP})), SP \cite{Dai_Subspace_pursuit} and CoSaMP \cite{Needell_Tropp_CoSaMP}. 

In contrast to the OMP algorithm (see Appendix~\ref{subsec:Serial_Atom_Selection_Based_OMP_and_OLS} for a brief description), the new additions in the POMP algorithm are the use of the set of potential atoms and the orthogonal projection based atom selection strategy from the set. 
A natural question is why does the projection based atom selection method perform better than the standard matched filter. For understanding, let us consider an over-determined case where $M \geq N$ and assume that the columns of $\mathbf{A}$ (i.e. atoms) are orthonormal \footnote{On this assumption, we mention that the construction of $\mathbf{A}$ by using i.i.d random variables guarantees a near-orthonormal condition by high probability.}. Now let us consider OMP for such an overdetermined $\mathbf{A}$ and assume clean measurement condition. In that case, the matched filter provides exact signal component values and hence the use of matched filter is optimal. The matched filter selects all the underlying $K$ atoms perfectly one-by-one over iterations. The selection order of atoms will follow a strictly regular path as follows: the atom associated with largest amplitude scalar value will be selected first and then the atom associated with second largest amplitude scalar value will be selected, and so on. That means the atom selection order will follow the strict path associated with the decreasing amplitudes of underlying non-zero scalar signal values. Now let us consider the underdetermined CS setup (or sparse signal estimation problem) where $M < N$ and hence all the atoms can not be orthonormal. In that case, the matched filter can not provide exact signal component values, but a mixed output and hence the use of matched filter is non-optimal. For any chosen atom, there exists a set of atoms which are highly correlated with the chosen atom; this set of atoms are coherent with the chosen atom. The mixed output of the matched filter is the outcome of the fact that an underlying atom is subjected to an interference from other atoms, mainly from the coherent atoms. The interference is due to non-orthogonality and hence the inner product operations in the matched filter do not provide reliable signal component values. For example, the highest amplitude coefficient of the output vector of the matched filter may not be associated with the corresponding underlying atom. Thus, the highest amplitude based detection strategy in the matched filter may not provide perfect selection of atoms over iterations. If $M \ll N$, the level of interference is considerably high and efficiency of the matched filter degrades severely, leading to poor atom selection. To circumvent the problem partially, we propose to use the orthogonal projection method in POMP. In each iteration of POMP, the orthogonal projection helps to estimate the signal component values associated with the chosen potential atoms in the LS sense. In general, the LS approach is powerful to estimate a signal. Thus, the LS signal estimation followed by the highest amplitude based atom selection strategy has a better potential than the sole use of matched filter. Note that the matched filter provides a LS solution for a single atom, without considering contributions from the non-orthogonal atoms.       

Next we discuss the complexity of the POMP algorithm where the orthogonal projection operation is performed twice in each iteration. Considering that the matched filtering and orthogonal projection operations are the most computationally intensive among all the relevant computational operations, the POMP complexity can be seen loosely as less than twice of the OMP complexity. We assume that the other computational operations require negligible resources. 
To be quantitative, the POMP requires to perform $K$ matched filter operations which is same as that of OMP. However, the POMP requires $2K$ orthogonal projection operations compared to the $K$ orthogonal projection operations in OMP.
In the POMP, we used the parameter $L$ (associated with the number of potential atoms) such that $L \leq K$. 
Our engineering perspective for the choice of $L \leq K$ is that no more atoms, than the total number of atoms that will eventually be chosen (i.e., $K$ atoms), need to be tested as the potential atoms in any iteration. Also, another consideration is that the atom selection performance saturates with increasing $L$.
Through experimental evaluations in section~\ref{subsec:Experimental_Results}, we show that an increase in $L$ leads to a better atom selection performance initially and then saturates. Observe that for $L=1$, the performance of POMP is the same as the OMP. 

In the POMP, the atom selection strategy for the current iteration does not depend on the future iterations. In the next section, we propose an algorithm that sees the future.

\section{Look Ahead Orthogonal Least Squares}
\label{sec:LAOLS}

In this section, we develop look ahead OLS (LAOLS) algorithm where the selection of an atom's index in the current iteration is carried out according to its future effect on minimizing the final residual norm. The final residual norm is evaluated at the end of all future iterations. For developing the LAOLS algorithm, we first need to realize a look ahead strategy in an algorithmic manner such that the future effect can be evaluated. Algorithm~\ref{alg:LA_strategy} shows such a look ahead strategy to evaluate the final residual norm.

\begin{algorithm}[ht!]
\caption{: Look Ahead Residual Norm}\label{alg:LA_strategy}
\mbox{Input: }
\begin{algorithmic}[1]
\STATE $\mathbf{A}$, $\mathbf{y}$, $K$;
\STATE Previous support set $\mathcal{I}$, current atom-index $i$;
\end{algorithmic}
\mbox{Assumption: }
\begin{algorithmic}[1]
\STATE $i \notin \mathcal{I}$;
\end{algorithmic}
\mbox{Initialization: }
\begin{algorithmic}[1]
\STATE Iteration counter $k\leftarrow| \mathcal{I} \cup i |$;
\STATE $\mathcal{I}_{k} \leftarrow \mathcal{I} \cup i$, $\mathbf{r}_{k}\leftarrow\mathbf{y} - \mathbf{A}_{\mathcal{I}_{k}} \, \mathbf{A}_{\mathcal{I}_{k}}^{\dag} \mathbf{y}$;
\end{algorithmic}
\mbox{Iterations: }
\begin{algorithmic}[1]
\REPEAT 
\STATE $k \leftarrow k+1$;
\STATE $i_{k} \leftarrow $ index of the highest amplitude of $\mathbf{A}^{t} \mathbf{r}_{k-1}$ \newline 
(Assumption: $i_{k} \notin \mathcal{I}_{k-1}$);
\label{step:atom_index_LA_strategy}
\STATE $\mathcal{I}_{k} \leftarrow \mathcal{I}_{k-1} \cup i_{k}$; \hfill (Note: $|\mathcal{I}_{k}|=k$) \label{step:Intermediate_Support_LA_strategy}
\STATE $\mathbf{r}_{k} \leftarrow \mathbf{y}-\mathbf{A}_{\mathcal{I}_{k}} \mathbf{A}_{\mathcal{I}_{k}}^{\dag} \mathbf{y}$; \hfill (Orthogonal projection)	\label{step:Proj_Residue_LA_strategy}
\UNTIL $( (\| \mathbf{r}_{k} \|_{2} > \| \mathbf{r}_{k-1} \|_{2}) \,\, \mathrm{or}\,\, (k>K) )$
\STATE $k \leftarrow k-1$; \hfill (Previous iteration)
\end{algorithmic}
\mbox{Output:}
\begin{algorithmic}[1]
\STATE $\| \mathbf{r}_{k} \|_{2}$ (where $\mathbf{r}_{k} = \mathbf{y}-\mathbf{A}_{\mathcal{I}_{k}} \mathbf{A}_{\mathcal{I}_{k}}^{\dag} \mathbf{y}$).
\end{algorithmic}
\end{algorithm}

The Algorithm~\ref{alg:LA_strategy} can be seen as a look ahead part of the OMP algorithm in the current iteration where the future atoms are chosen using the matched filter and orthogonal projection operations. Given the sparsity level $K$, a previous support set $\mathcal{I}$ and a current atom-index $i$ (index of the atom chosen in the current iteration), the algorithm usually finds a $K$-element support set at the end of all future iterations and returns the final LS residual norm. For the given previous support set $\mathcal{I}$, the final effect of the choice of a new atom in the current iteration can be evaluated through using this look ahead part. 
 
Using Algorithm~\ref{alg:LA_strategy}, let us define the following function.
\begin{myfunc}
\label{func:Look_ahead_residual_norm}
(Look ahead residual norm) Let $\mathbf{y} \in \mathbb{R}^{M}$, $\mathbf{A} \in \mathbb{R}^{M \times N}$ and the sparsity level $K$ are given. Suppose that the previous support set is $\mathcal{I}$ and the current atom-index is $i$. Then the look ahead residual norm $ n \in \mathbb{R}$ is the output of the following algorithmic function  
\begin{eqnarray}
n = \mathrm{look \_ ahead \_ resid \_ norm}\left(\mathbf{A}, \mathbf{y}, K, \mathcal{I}, i \right) 
\end{eqnarray}
where the above function executes the Algorithm~\ref{alg:LA_strategy}.
\end{myfunc}

Using Function~\ref{func:Look_ahead_residual_norm}, the main steps of the LAOLS algorithm are summarized in Algorithm~\ref{alg:LAOLS}.
\begin{algorithm}[ht!]
\caption{: LAOLS for CS Recovery}\label{alg:LAOLS}
\mbox{Input: }
\begin{algorithmic}[1]
\STATE $\mathbf{A}$, $\mathbf{y}$, $K$;
\STATE $L$ \hfill ($L$ potential atoms and $1 \leq L  \leq K$);
\end{algorithmic}
\mbox{Define: }
\begin{algorithmic}[1]
\STATE $\mathbf{j}=\left[ j_{1} \,\, j_{2}, \ldots, {j}_{L} \right]^{t}$, $\mathbf{n}=\left[ n_{1} \,\, n_{2}, \ldots, {n}_{L} \right]^{t}$;
\end{algorithmic}
\mbox{Initialization: }
\begin{algorithmic}[1]
\STATE Iteration counters $k \leftarrow 0$ (Outer loop), $l \leftarrow 0$ (Inner loop);
\STATE $\mathbf{r}_{0} \leftarrow \mathbf{y}$, $\mathcal{I}_{0} \leftarrow \emptyset$;
\end{algorithmic}
\mbox{Iterations: }
\begin{algorithmic}[1]
\REPEAT 
\STATE $k \leftarrow k+1$;
\STATE $\mathcal{I}_{(p)} \leftarrow$ $\{$indices of $L$ highest amplitudes of $\mathbf{A}^{t} \mathbf{r}_{k-1}\}$; \newline
(Assumption: chosen $L$ indices $\notin \mathcal{I}_{k-1}$) \label{step:Set_of_Potential_Atoms_LAOLS}
\STATE $\mathbf{j} \leftarrow \mathcal{I}_{(p)}$; \newline (Elements of $\mathcal{I}_{(p)}$ are assigned to components of $\mathbf{j}$)
\FOR{$l=1$ to $L$}
  \STATE $n_{l} \leftarrow \mathrm{look \_ ahead \_ resid \_norm}\left( \mathbf{A}, \mathbf{y}, K, \mathcal{I}_{k-1}, j_{l} \right)$; \newline
  (Final residual norm if $j_{l}$ is selected)
  \ENDFOR
\STATE $l^{\star} \leftarrow $ coordinate of the lowest component of $\mathbf{n}$; \label{step:Lowest_residual_norm_LAOLS}
\STATE $i_{k} \leftarrow j_{l^{\star}}$; \newline 
(Index of atom providing lowest final residual norm) 
\label{step:Selected_atom_index_LAOLS}
\STATE $\mathcal{I}_{k} \leftarrow \mathcal{I}_{k-1} \cup i_{k}$; \hfill (Note: $|\mathcal{I}_{k}|=k$) \label{step:Intermediate_Support_LAOLS}
\STATE $\mathbf{r}_{k} \leftarrow \mathbf{y}-\mathbf{A}_{\mathcal{I}_{k}} \mathbf{A}_{\mathcal{I}_{k}}^{\dag} \mathbf{y}$; \hfill (Orthogonal projection)	\label{step:Proj_Residue_LAOLS}
\UNTIL $( (\| \mathbf{r}_{k} \|_{2} > \| \mathbf{r}_{k-1} \|_{2}) \,\, \mathrm{or}\,\, (k>K) )$
\STATE $k \leftarrow k-1$; \hfill (Previous iteration)
\end{algorithmic}
\mbox{Output:}
\begin{algorithmic}[1]
\STATE $\hat{\mathbf{x}} \in \mathbb{R}^{N}$, satisfying $\hat{\mathbf{x}}_{\mathcal{I}_{k}} = \mathbf{A}_{\mathcal{I}_{k}}^{\dag} \mathbf{y}$ and $\hat{\mathbf{x}}_{\overline{\mathcal{I}}_{k}} = \mathbf{0}$.
\end{algorithmic}
\end{algorithm}
Like POMP, we fix an integer parameter $L \leq K$ that decides how many potential atoms are checked in each iteration using the look ahead strategy of Function~\ref{func:Look_ahead_residual_norm} (or Algorithm~\ref{alg:LA_strategy}).    
The atom with smallest look ahead residual norm is selected and added to the previous support set $\mathcal{I}_{k-1}$ to form the new support set $\mathcal{I}_{k}$ of cardinality $k$. 
Given the sparsity level $K$, the algorithm usually executes $K$ iterations and forms a support set of cardinality $K$ at the end.
We mention that the LAOLS algorithm can be seen as a confluence of OMP and OLS algorithms. For the OMP connection, note the use of matched filter in the look ahead  part to find the indices of future atoms (see step~\ref{step:atom_index_LA_strategy} of Algorithm~\ref{alg:LA_strategy}). Also, for the OLS connection, note how the index of a new atom is selected in each iteration depending on the minimum norm of fitting residual (see steps~\ref{step:Lowest_residual_norm_LAOLS} and~\ref{step:Selected_atom_index_LAOLS} of Algorithm~\ref{alg:LAOLS}). Technically, for designing a look ahead OLS, it is possible to use the OLS strategy of atom selection in Algorithm~\ref{alg:LA_strategy} instead of using the OMP strategy. However, we refrain from such a strategy due to a heavy increase in computational complexity\footnote{The atom selection strategy of OLS is more computationally intensive than the atom selection strategy of OMP. The reader is encouraged to see Appendix~\ref{subsec:Serial_Atom_Selection_Based_OMP_and_OLS}.}.

In contrast to the OLS algorithm (see Appendix~\ref{subsec:Serial_Atom_Selection_Based_OMP_and_OLS} for a brief description), the new additions in the LAOLS algorithm are the use of a set of potential atoms and the look ahead atom selection from the set. A natural observation is that the use of look ahead strategy is computationally intensive. 
To achieve computational advantage, the orthogonal projection operations in the look ahead strategy can be performed recursively using block-wise matrix inversion. 
An exposition for performing recursive computation is shown in Appendix~\ref{subsec:Resursive_Computation_for_Look_Ahead}. For each $l$ at the $k$th iteration, the look ahead strategy requires to perform $(K-(k-1))$ matched filtering and orthogonal projection operations. Therefore, each of the total matched filtering and the total orthogonal projection operations in the LAOLS are $\approx \sum_{l=1}^{L} \sum_{k=1}^{K} L (K-(k-1)) = \sum_{k=1}^{K} L (K-(k-1)) = L(1+2+ \ldots +K) = \frac{1}{2}L(K^2+K)$. Given the sparsity level $K$, the computational complexity increases linearly with the number of potential atoms $L$. A reduction in $L$ leads to a lower complexity, but at an expense of CS recovery performance. In the next section, we develop an algorithm with a reduced complexity, but without much loss in performance.

\section{Structured Orthogonal Least Squares}
\label{sec:Structured_Orthogonal_Least_Squares}

In this section, we develop a structured IGS algorithm by combining POMP and LAOLS algorithms such that a trade-off between computational complexity and reconstruction performance can be established. In each iteration, we choose $L$ potential atoms by the standard matched filter followed by reducing the cardinality of the set of potential atoms through an orthogonal projection based atom selection strategy. Then, we use the look ahead strategy to select an atom from the reduced set of potential atoms. The use of the reduced set allows for a sharp reduction in computational burden. However, as the cardinality reduction of the set of potential atoms is performed through an orthogonal projection method, a little performance loss can be expected. Note that an atom is finally selected from the set of $L$ potential atoms through using the two selection schemes which are algorithmically in a cascade connection. The new algorithm is referred to as structured OLS (SOLS). 

To develop the SOLS algorithm, we first need to describe the Algorithm~\ref{alg:Proj_Multiple_atom_selection} which is a slight modification of Algorithm~\ref{alg:Proj_atom_selection}.     
\begin{algorithm}[ht!]
\caption{: Projection-Based Multiple Atom-indices Selection}\label{alg:Proj_Multiple_atom_selection}
\mbox{Input: }
\begin{algorithmic}[1]
\STATE $\mathbf{A}$, $\mathbf{y}$;
\STATE Previous support set: $\mathcal{I}$;
\STATE Set of potential atoms' indices: $\mathcal{I}_{(p)}$; 
\STATE $L^{\prime}$ ($L^{\prime} \leq |\mathcal{I}_{(p)}|$);
\end{algorithmic}
\mbox{Assumption: }
\begin{algorithmic}[1]
\STATE $\mathcal{I}_{(p)} \cap \mathcal{I} = \emptyset$, $\hat{\mathbf{x}} \in \mathbb{R}^{N}$;
\end{algorithmic}
\mbox{Execution: }
\begin{algorithmic}[1]
\STATE $\mathcal{I}_{(u)} \leftarrow \mathcal{I} \cup \mathcal{I}_{(p)}$;
\STATE $\hat{\mathbf{x}}_{\mathcal{I}_{(u)}} \leftarrow \mathbf{A}_{\mathcal{I}_{(u)}}^{\dag} \mathbf{y}$; \hfill (Orthogonal projection)
\STATE $\hat{\mathbf{x}}_{\overline{\mathcal{I}}_{(p)}} \leftarrow \mathbf{0}$; \hfill (Zeroing the coordinates indexed by $\overline{\mathcal{I}}_{(p)}$)
\STATE $\mathcal{I}_{(p^{\prime})} \leftarrow \mathrm{indices \,\, of \,\, the \,\, L^{\prime} \,\, highest \,\, amplitudes \,\, of \,\, \hat{\mathbf{x}}}$;
\end{algorithmic}
\mbox{Output:}
\begin{algorithmic}[1]
\STATE The indices' set $\mathcal{I}_{(p^{\prime})}$.
\end{algorithmic}
\end{algorithm}
In the Algorithm~\ref{alg:Proj_Multiple_atom_selection}, $L^{\prime}$ is a positive integer that decides how many atoms are selected from the set of potential atoms $\mathcal{I}_{(p)}$ through orthogonal projection. Considering $L^{\prime} \leq |\mathcal{I}_{(p)}|$, the set of $L^{\prime}$ atoms are used for further processing. For a given cardinality of the set $\mathcal{I}_{(p)}$, the choice of $L^{\prime}$ is again a user's prerogative that depends on the trade-off between atom selection precision and search complexity. 

Using Algorithm~\ref{alg:Proj_Multiple_atom_selection}, let us define the following function.
\begin{myfunc}
\label{func:Projection_based_multiple_atom-indices_selection}
(Projection-based multiple atom-indices selection) Let $\mathbf{y} \in \mathbb{R}^{M}$, $\mathbf{A} \in \mathbb{R}^{M \times N}$, previous support set $\mathcal{I}$ and the set of potential atoms' indices $\mathcal{I}_{(p)}$ are given. Suppose $\mathcal{I}_{(p^{\prime})} \subset \mathcal{I}_{(p)}$ denotes the set of indices of $L^{\prime}$ atoms selected in the current iteration. Then $\mathcal{I}_{(p^{\prime})}$ is the output of the following algorithmic function  
\begin{eqnarray}
\mathcal{I}_{(p^{\prime})} = \mathrm{proj \_ multi \_ atom \_ indices}\left( \mathbf{A}, \mathbf{y}, \mathcal{I}, \mathcal{I}_{(p)}, L^{\prime} \right) 
\end{eqnarray}
where the above function executes the Algorithm~\ref{alg:Proj_Multiple_atom_selection}.
\end{myfunc}

Using Functions~\ref{func:Look_ahead_residual_norm} and~\ref{func:Projection_based_multiple_atom-indices_selection}, the main steps of the SOLS algorithm are summarized in Algorithm~\ref{alg:SOLS}.
\begin{algorithm}[ht!]
\caption{: SOLS for CS Recovery}\label{alg:SOLS}
\mbox{Input: }
\begin{algorithmic}[1]
\STATE $\mathbf{A}$, $\mathbf{y}$, $K$;
\STATE $L$ \hfill ($L$ potential atoms and $1 \leq L \leq K$);
\STATE $\gamma$ \hfill (A subset selection parameter and $0 \leq \gamma < 1$). 
\end{algorithmic}
\mbox{Define: }
\begin{algorithmic}[1]
\STATE $L^{\prime} = L - \lfloor \gamma L \rfloor$; \hfill ($L^{\prime} > 0$; non-zero cardinality of subset)
\STATE $\mathbf{j}=\left[ j_{1} \,\, j_{2}, \ldots, {j}_{L^{\prime}} \right]^{t}$, $\mathbf{n}=\left[ n_{1} \,\, n_{2}, \ldots, {n}_{L^{\prime}} \right]^{t}$;
\end{algorithmic}
\mbox{Initialization: }
\begin{algorithmic}[1]
\STATE Iteration counters $k \leftarrow 0$ (Outer loop), $l \leftarrow 0$ (Inner loop);
\STATE $\mathbf{r}_{0} \leftarrow \mathbf{y}$, $\mathcal{I}_{0} \leftarrow \emptyset$;
\end{algorithmic}
\mbox{Iteration: }
\begin{algorithmic}[1]
\REPEAT
\STATE $k \leftarrow k+1$;
\STATE $\mathcal{I}_{(p)} \leftarrow $ indices of the $L$ highest amplitudes of $\mathbf{A}^{t} \mathbf{r}_{k-1}$; \newline
(Assumption: chosen $L$ indices $\notin \mathcal{I}_{k-1}$) 
\STATE $\mathcal{I}_{(p^{\prime})} \leftarrow $ $\mathrm{proj \_ multi \_ atom \_ indices}\left(\mathbf{A}, \mathbf{y}, \mathcal{I}_{k-1}, \mathcal{I}_{(p)}, L^{\prime} \right)$; \newline
(Reducing the number of potential atoms: $| \mathcal{I}_{(p^{\prime})}|=L^{\prime}$) \label{step:Proj_atoms_indices_SOLS}
\STATE $\mathbf{j} \leftarrow \mathcal{I}_{(p^{\prime})}$; \newline (Elements of $\mathcal{I}_{(p^{\prime})}$ are assigned to components of $\mathbf{j}$)
  \FOR{$l=1$ to $L^{\prime}$}
  \STATE $n_{l} \leftarrow \mathrm{look \_ ahead \_ resid \_norm}\left( \mathbf{A}, \mathbf{y}, K, \mathcal{I}_{k-1}, j_{l} \right)$;
  \ENDFOR
\STATE{$l^{\star} \leftarrow \mathrm{coordinate \,\, of \,\, the \,\, lowest \,\, component \,\, of \,\, \mathbf{n}}$;} 
\STATE{$i_{k} \leftarrow j_{l^{\star}}$; \newline (Index of the atom providing lowest final residual norm)}
\STATE{$\mathcal{I}_{k} \leftarrow \mathcal{I}_{k-1} \cup i_{k}$;}
\STATE $\mathbf{r}_{k} \leftarrow \mathbf{y}-\mathbf{A}_{\mathcal{I}_{k}} \mathbf{A}_{\mathcal{I}_{k}}^{\dag} \mathbf{y}$; \hfill (Orthogonal projection)	\label{step:Proj_Residue_SOLS}
\UNTIL{$( (\| \mathbf{r}_{k} \|_{2} > \| \mathbf{r}_{k-1} \|_{2}) \,\, \mathrm{or}\,\, (k>K) )$}
\STATE {$k \leftarrow k-1$; \hfill (Previous iteration count)}
\end{algorithmic}
\mbox{Output:}
\begin{algorithmic}[1]
\STATE $\hat{\mathbf{x}} \in \mathbb{R}^{N}$, satisfying $\hat{\mathbf{x}}_{\mathcal{I}_{k}} = \mathbf{A}_{\mathcal{I}_{k}}^{\dagger} \mathbf{y}$ and $\hat{\mathbf{x}}_{\overline{\mathcal{I}}_{k}} = \mathbf{0}$.
\end{algorithmic}
\end{algorithm}
In the algorithm, we introduce an input parameter $\gamma$ ($0 \leq \gamma < 1$) that decides the cardinality of reduced set of potential atoms as $L^{\prime} = L - \lfloor \gamma L \rfloor$. In the iterations, the selection of atoms through orthogonal projection is performed in step~\ref{step:Proj_atoms_indices_SOLS} where $L^{\prime}$ atoms are chosen from the set of $L$ potential atoms. Then the chosen $L^{\prime}$ atoms are subjected through look ahead strategy such that a single atom is finally selected from the $L^{\prime}$ atoms and added to the intermediate support-set. For $\gamma = 0$, SOLS acts as LAOLS; on the other hand, for a $\gamma$ near to one that results in $L^{\prime}=1$, SOLS acts as POMP. 

Each of the total matched filtering and the total orthogonal projection operations in SOLS $\approx \sum_{k=1}^{K} \left[ 1+ L^{\prime} (K-(k-1)) \right] = K + \frac{1}{2} L^{\prime} (K^2+K) = K + \frac{1}{2} (L - \lfloor \gamma L \rfloor) (K^2+K)$. Given the sparsity level $K$ and the number of potential atoms $L$, the computational complexity decreases linearly with the increase in $\gamma$. For example, the choice of $\gamma=0.5$ leads to a reduction of computational complexity by nearly half compared to the LAOLS algorithm.

\section{Comparison of Computational Complexity}
\label{sec:Comparison_of_Computational_Complexity}

In this section, we perform an analysis of computational complexity of the new algorithms and compare them with existing algorithms. We mainly quantify the number of addition and multiplication operations. 

Let us first consider the POMP algorithm. For $k$'th iteration, the algorithm requires one matched filter operation and two orthogonal projections. A matched filter operation requires $2MN$ computations and each orthogonal projection requires $\mathcal{O}(k^2M)$ computations. Overall the associated complexity is $\mathcal{O}(MN+K^2M)$ where we assume the worst case scenario that $k=K$. 
Now, considering $K$ iterations, the overall complexity of the POMP algorithm is $\mathcal{O}(K(MN+K^2M))$.

Next we consider the LAOLS algorithm. In this case, the intensive computation is due to the look ahead part where it is necessary to perform matched filter and orthogonal projection operations. We compute orthogonal projections recursively using the block-wise matrix inversion mechanism shown in Appendix~\ref{subsec:Resursive_Computation_for_Look_Ahead}. Using recursive computation and considering the worst case scenario that $k=K$, the necessary computation for each orthogonal projection is approximately $7K^2+4KM$. So, the total computation for each matched filter and each orthogonal projection is $\mathcal{O}(MN+K^2+KM)$. 
Now, considering $\frac{1}{2}L(K^2+K)$ matched filter operations and orthogonal projections, the total complexity of LAOLS is $\mathcal{O}(LK^2(MN+K^2+KM))$.
Following the similar arguments, the complexity of SOLS algorithm is $\mathcal{O}((L - \lfloor \gamma L \rfloor)K^2(MN+K^2+ KM ))$. 

\footnotesize
\renewcommand{\baselinestretch}{1}
\begin{table}[t]
\centering
\caption{Computational complexity comparison}
\setlength{\tabcolsep}{4pt}

\begin{tabular}{c||c}
	\hline

	\multicolumn{1}{c||}{Algorithm}	&
	\multicolumn{1}{c}{Order of complexity}	\\ \hline \hline

	\multicolumn{2}{c}{Existing algorithms}	\\ \hline \hline

	OMP    & $\mathcal{O}(K(MN+K^2+KM))$    	\\
	OLS    & $\mathcal{O}(K^2(KN+MN))$      	\\
	SP     & $\mathcal{O}(K(MN + K^2M))$			\\
        Convex relaxation & $\mathcal{O}(N^2M)$ or $\mathcal{O}(N^3)$ 	\\  	\hline 

	\multicolumn{2}{c}{New algorithms}	\\ \hline \hline
        POMP   & $\mathcal{O}(K(MN+K^2M))$	\\
	LAOLS  & $\mathcal{O}(LK^2(MN+K^2+KM))$	\\
	SOLS  & $\mathcal{O}((L - \lfloor \gamma L \rfloor)K^2(MN+K^2+KM ))$	\\    \hline
        
	\end{tabular}
	
	\label{table:ComputationalComplexityComparison_of_POMP_LAOLS}
	
\end{table}
\renewcommand{\baselinestretch}{1}
\normalsize 

We show the comparison of complexity between several algorithms in Table~\ref{table:ComputationalComplexityComparison_of_POMP_LAOLS}. Following \cite{Boyd_Book} (pages 6-8 of \cite{Boyd_Book}), we assume that complexity of convex relaxation methods is either $\mathcal{O}(N^2M)$ (using linear program) or $\mathcal{O}(N^3)$ (using interior-point methods). The complexity of the existing IGS algorithms are evaluated in Appendix~\ref{subsec:Serial_Atom_Selection_Based_OMP_and_OLS} and \ref{subsec:Parallel_Atom_Selection_Based_SP}. Let us consider a quantitative analysis to show the complexity benefit of LAOLS over convex relaxation methods. From the literature, we note that most of the sparse reconstruction algorithms are successful when $M=\mathcal{O}(K \log N)$. For our comparison, we assume that $K$ varies linearly with $N$ as $K=\tau N$ (where $0 < \tau < 1$) and a typical setup where $K = \tau N \ll M=K \log N \ll N$. We also use $L=K$ for LAOLS and find its computational complexity as $\mathcal{O}(K^3MN)$; the term $K^2+KM$ is neglected as $K^2+KM \ll MN$. Therefore, the computational complexity of LAOLS is $\mathcal{O}(K^3MN) = \mathcal{O}(K^4N \log N) = \mathcal{O}(\tau^4 N^5 \log N)$. We also assume that interior-point methods are used for convex relaxation algorithms and hence their complexity is $\mathcal{O}(N^3)$. To achieve a better computational advantage for LAOLS than the convex relaxation algorithms, we must need $\mathcal{O}(\tau^4 N^5 \log N) < \mathcal{O}(N^3)$. This may be satisfied when $\tau^4 N^5 \log N < N^3$, or $\tau < (N^2 \log N)^{-0.25}$. For example, when $N=500$, we need $\tau < 0.0283$ and hence the allowed $K < 0.0283 \times 500 \approx 15$. The above analysis is also valid for SOLS. 

To achieve complexity advantage for LAOLS/SOLS over convex relaxation algorithms, the above analysis shows the need of a lower $K$.
The complexity increases heavily for a higher $K$ and hence, it is natural to seek suboptimal solutions for a higher $K$. A possible suboptimal solution may be not to allow the full future evolution in look ahead strategy, but to allow some extent by restricting the number of future iterations. Considering a depth variable $\eta$ that denotes the necessary allowable future iterations performed in look ahead strategy, a trade-off between performance and computational complexity may be achieved by proper choice of $L$ and $\eta$. An engineering intuition is that the increase of any of them keeping the other fixed leads to a saturation in performance. In the next section, we allow the look ahead strategy to check the full future evolution and show saturation in performance for increasing $L$. We also show running time comparison results between competing algorithms.

\section{Experiments and Results}

We performed computer simulations in order to compare the performance of following CS reconstruction algorithms: OMP, OLS, SP, POMP, LAOLS, SOLS, and basis pursuit/basis pursuit denoising/LASSO (BP/BPDN/LASSO) \cite{CS_introduction_Candes_Wakin}-{\cite{Chen_Donoho_Saunders}, \cite{Tibshirani_lasso}-\cite{Efron_LARS}. Among these algorithms, BP/BPDN/LASSO is a convex relaxation algorithm ($l_{1}$ norm minimization based) and the other algorithms are IGS methods. The BP/BPDN simulation code (matlab) was taken from the $l_{1}$-magic toolbox \cite{Candes_Romberg_L1_Magic_Toolbox}. The LASSO simulation code (matlab) was taken from the SparseLab toolbox \cite{SparseLab_Toolbox}. For SOLS, we used a fixed value of $\gamma = 0.5$. Experimentally we evaluated the effect of $L$ on the trade-off between complexity and reconstruction performance for POMP/LAOLS/SOLS.  We first discuss the reconstruction performance measures and experimental setups, and then discuss the BP/BPDN/LASSO algorithm briefly followed by reporting the performance results of all the algorithms for clean and noisy measurements.

\subsection{Performance measures and experimental setups}

We use two performance measures. For the first performance measure, we use signal-to-reconstruction-noise ratio (SRNR) defined as
\begin{eqnarray}
\mathrm{SRNR} = \frac{ \mathcal{E} \{ \| \mathbf{x} \|_{2}^{2} \} }{\mathcal{E} \{ \| \mathbf{x} - \hat{\mathbf{x}} \|_{2}^{2} \} }, 
\end{eqnarray} 
where $\hat{\mathbf{x}}$ is the reconstructed signal vector. Note that our objective is to achieve a higher SRNR.

Next we define another performance measure which provides a direct measure of estimating the underlying support set. 
For a $K$-sparse signal vector $\mathbf{x}$, the support set was denoted as $\mathcal{I}$ with cardinality $K$. Let us denote the support set of reconstructed vector $\hat{\mathbf{x}}$ as $\hat{\mathcal{I}}$. We assume that $\hat{\mathbf{x}}$ is also a $K$-sparse signal vector, i.e. $|\hat{\mathcal{I}}|=K$.
To measure the support set estimation error, we consider to use the distortion $d(\mathcal{I},\hat{\mathcal{I}})= 1- \left( | \mathcal{I} \cap \hat{\mathcal{I}}| / K  \right)$ \cite{Reeves_Gastpar_Asilomar}. Considering a large number of realizations (signal vectors), we can compute the average of $d(\mathcal{I},\hat{\mathcal{I}})$. We define the average support-cardinality error (ASCE) as follows
\begin{eqnarray}
\mathrm{ASCE} = \mathcal{E} \left\{ d(\mathcal{I},\hat{\mathcal{I}}) \right\} = 1 - \frac{1}{K} \mathcal{E} \left\{ | \mathcal{I} \cap \hat{\mathcal{I}} | \right\}.
\label{eq:ASCE}
\end{eqnarray} 
Note that the ASCE has the range $[0, 1]$ and our objective is to achieve a lower ASCE. Along-with SRNR, the ASCE is used as the second performance evaluation measure because the main objective of the IGS algorithms is to estimate the underlying support set. 

Now we discuss experimental setups. In a CS setup, all sparse signal vectors are expected to be exactly reconstructed if the number of measurements is more than a certain threshold \cite{Dai_Subspace_pursuit}. However, the computational complexity to test this uniform reconstruction ability is very high. Instead, for empirical testing, we can devise a strategy that can compute the performance measures for random measurement matrix ensemble. To measure the level of under-sampling, let us define the fraction of measurements (FoM) 
\begin{eqnarray}
\alpha=\frac{M}{N}.
\end{eqnarray} 
Using $\alpha$, steps of the testing strategy are listed as follows:
\begin{enumerate}
\item {For given values of the parameters $K$ and $N$, choose $\alpha$ such that the number of measurements $M$ is an integer.} 
\item {Randomly generate a sensing matrix $\mathbf{A} \in \mathbb{R}^{M \times N}$ where the components are drawn independently from a Gaussian source (i.e., $a_{i,j} \sim \mathcal{N} \left(0, \frac{1}{M} \right)$) and then scale the columns of $\mathbf{A}$ to unit-norm.}
\item {Randomly generate a set of $K$-sparse data $\mathbf{x}$ where the support set $\mathcal{I}$ is chosen uniformly over the set $\{1,2,\ldots,N\}$. Let we denote the size of data as $S$ (i.e., the number of signal vectors $\mathbf{x}$ is $S$). The non-zero components of $\mathbf{x}$ are independently drawn by either of the following two methods. 
\begin{enumerate}
\item {The non-zero components are drawn independently from a standard Gaussian source. This type of signal is referred to as Gaussian sparse signal.}
\item {The non-zero components are set to ones. This type of signal is referred to as binary sparse signal.}
\end{enumerate}
Note that the Gaussian sparse signal is compressible in nature. That means, in the descending order, the sorted amplitudes of a Gaussian sparse signal vector's components decay fast with respect to the sorted indices. This decaying trend corroborates with several natural signals (for example, wavelet coefficients of an image). On the other hand, a binary sparse signal is not compressible in nature, but of special interest for comparative study, since it represents a particularly challenging case for OMP-type of reconstruction strategies \cite{Tropp_OMP}, \cite{Dai_Subspace_pursuit}.  
}
\item {For each data, compute the measurement $\mathbf{y} = \mathbf{A} \mathbf{x} + \mathbf{w}$ and apply the CS reconstruction methods independently.}
\item {Repeat steps 2-4 for a given times (let $T$ times). Then evaluate the CS performance evaluation measures (by averaging over $ST$ data).}
\item {Repeat steps 1-5 for a new $\alpha$.}
\end{enumerate}
This test can be performed for any chosen $K$ and $N$.

Considering the measurement noise $\mathbf{w} \sim \mathcal{N} \left( \mathbf{0}, \sigma_{w}^{2} \mathbf{I}_{M} \right)$, we define the signal-to-measurement-noise-ratio (SMNR) as
\begin{eqnarray}
\mathrm{SMNR} = \frac{ \mathcal{E} \{ \| \mathbf{x} \|_{2}^{2} \} }{ \mathcal{E} \{ \| \mathbf{w} \|_{2}^{2} \} } , 
\end{eqnarray} 
where $\mathcal{E} \{ \| \mathbf{w} \|_{2}^{2} \} = \sigma_{w}^{2}M$. We report the experimental results at varying SMNRs. 

In the presence of a measurement noise, it is impossible to achieve perfect CS recovery. On the other hand, for the clean measurement case, perfect CS recovery of a sparse signal is possible if $\alpha$ exceeds a certain threshold. In the spirit of using CS for practical applications with a less number of measurements at clean and noisy conditions, we are mainly interested in a lower range of $\alpha$ where performances of the contesting algorithms can be fairly compared.

\begin{figure*}
\centering
\includegraphics[width=7.3in,height=1.6in]{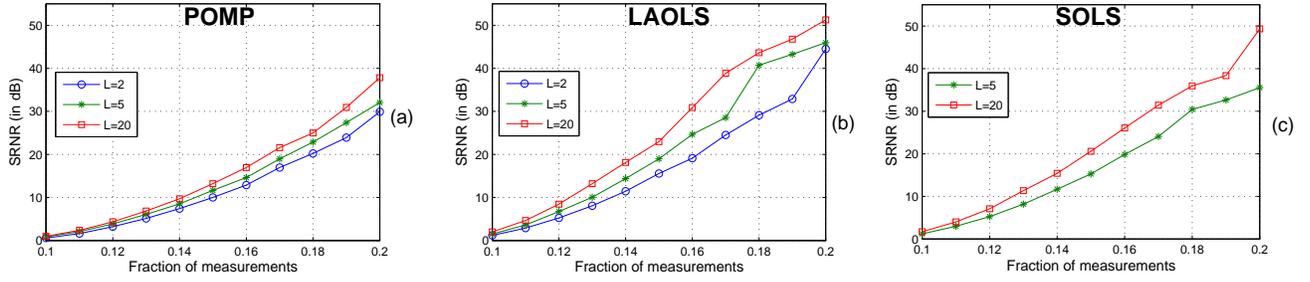}
\vspace{-2mm}
\caption{Effect of increasing $L$ on the POMP, LAOLS and SOLS algorithms for Gaussian sparse signal at the clean measurement condition. SRNR (in dB) versus fraction of measurements ($\alpha$) is plotted: (a) POMP, (b) LAOLS, and (c) SOLS.}
\label{fig:Effect_of_L_on_AllOMPs_Gaussian_CleanMeasurement_1}
\vspace{-3mm}
\end{figure*}

\begin{figure*}
\centering
\includegraphics[width=10in,height=6.2in,angle=90]{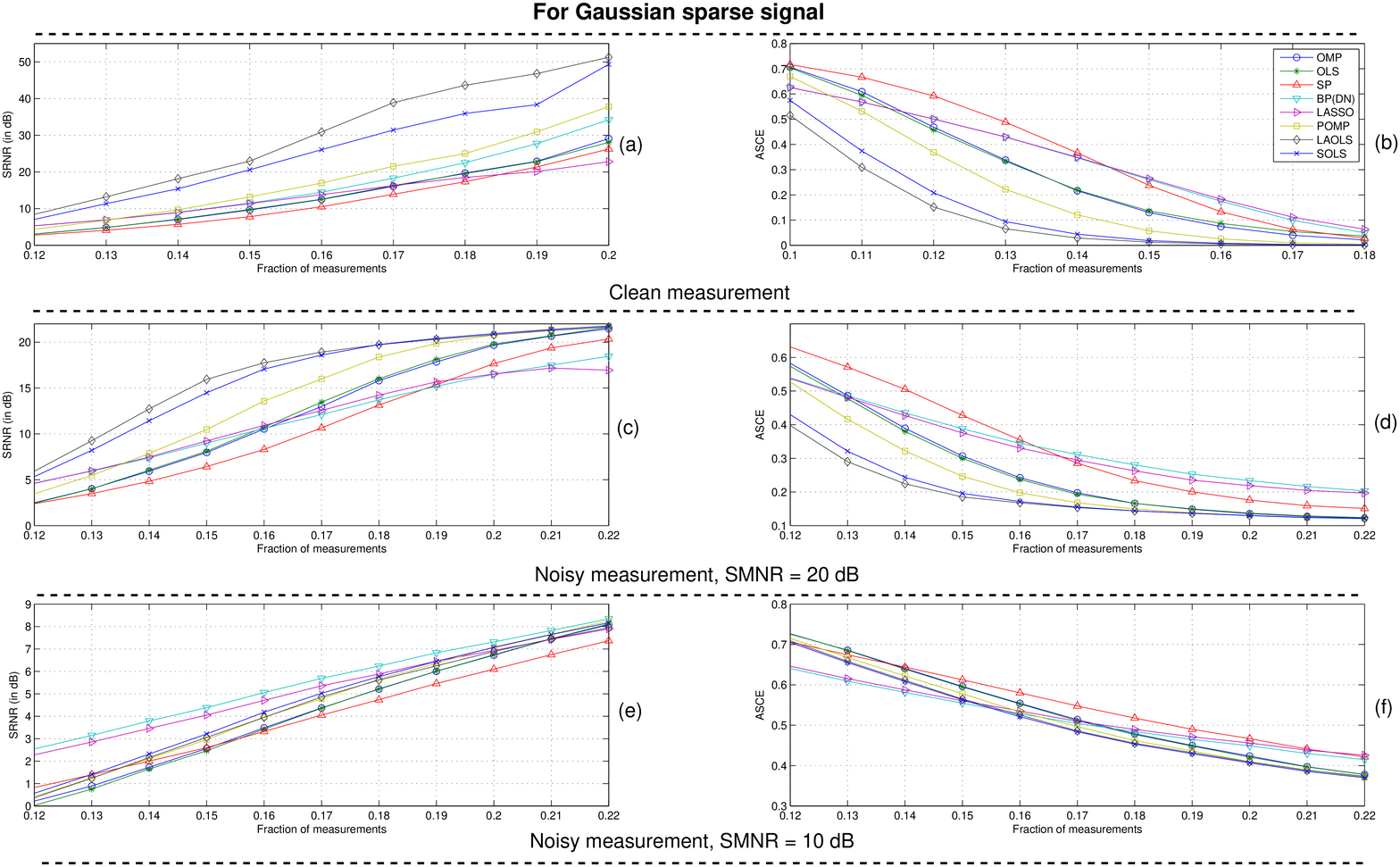}
\caption{Comparison of OMP, OLS, SP, BP/BPDN, LASSO, POMP, LAOLS and SOLS algorithms for {\bf{Gaussian sparse signal in clean and noisy measurement conditions}}. SRNR (in dB) and ASCE performances are plotted against the fraction of measurements ($\alpha$). We show the results for clean measurement as well as noisy measurements with SMNR = 20 and 10 dB. BP/BPDN is denoted by BP(DN).}
\label{fig:Comparison_Using_Gaussian_Signal_3}
\end{figure*}

\begin{figure*}
\centering
\includegraphics[width=5.5in,height=3.7in]{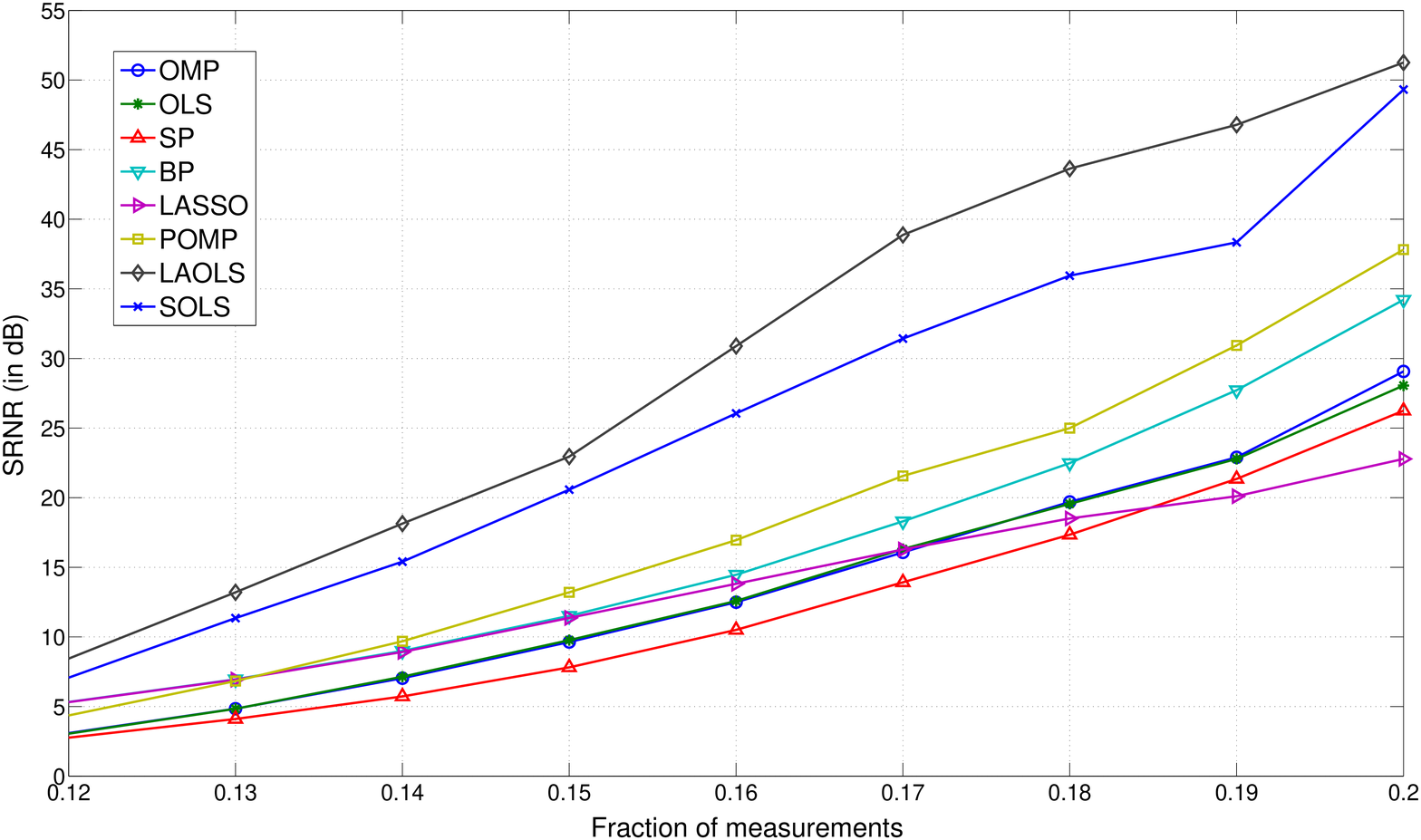}
\caption{Comparison of OMP, OLS, SP, BP, LASSO, POMP, LAOLS and SOLS algorithms for {\bf{Gaussian sparse signal in clean measurement condition}}. SRNR (in dB) is plotted against the fraction of measurements ($\alpha$). The same results are shown in Fig.~\ref{fig:Comparison_Using_Gaussian_Signal_3} (a).}
\label{fig:Comparison_Using_Gaussian_Signal_3_SRNR_CleanMeasurement}
\end{figure*}

\subsection{BP/BPDN/LASSO} 

In this subsection, we discuss BP, BPDN and LASSO \cite{CS_introduction_Candes_Wakin}. The BP/BPDN is a convex relaxation algorithm that looks for directly minimizing the $l_{1}$ norm of the solution vector with respect to convex constraint functions. For the case of clean measurement (i.e. $\mathbf{w}=0$ in (\ref{eq:CS_Standard_Problem})), we use BP where the following optimization problem is solved
 \begin{eqnarray}
\def\argmin{\mathop{\rm{arg\hspace{2pt}min}}}
P_{1}: \hspace{0.5cm}\hat{\mathbf{x}} = \argmin_{\mathbf{x} \in \mathbb{R}^{N}} \hspace{5pt}
\| \mathbf{x} \|_{1} \hspace{5pt} \mathrm{subject} \hspace{3pt} \mathrm{to} \hspace{3pt} 
\mathbf{y} = \mathbf{A} \mathbf{x}. 
\label{eq:CS_Standard_Problem_BP}
\end{eqnarray} 
The above optimization problem can be recast as a linear program (LP) \cite{Candes_Romberg_L1_Magic_Toolbox}. For the case of noisy measurements, we use BPDN \cite{CS_introduction_Candes_Wakin}. Using a quadratic constraint, we solve for 
\begin{eqnarray}
\def\argmin{\mathop{\rm{arg\hspace{2pt}min}}}
P_{1}^{\epsilon}: \hspace{0.5cm}\hat{\mathbf{x}} = \argmin_{\mathbf{x} \in \mathbb{R}^{N}} \hspace{5pt}
\| \mathbf{x} \|_{1} \hspace{5pt} \mathrm{subject} \hspace{3pt} \mathrm{to} \hspace{3pt} 
\| \mathbf{y} -  \mathbf{A} \mathbf{x} \|_{2} \leq \epsilon ,
\label{eq:CS_Standard_Problem_BPDN_CandesTheo}
\end{eqnarray}
where $\epsilon \geq \| \mathbf{w} \|_{2}$ is a parameter that depends on the noise properties. The above problem can be recast as a second order cone program (SOCP) \cite{Candes_Romberg_L1_Magic_Toolbox} and solved by an interior-point method. An interior-point method, such as the log-barrier method \cite{Boyd_Book}, is powerful and provides consistent performance. According to the suggestion of \cite{Candes_Romberg_Tao_Stable_recovery}, we choose $\epsilon^{2} = \sigma_{w}^{2} \left( M + 2 \sqrt{2M} \right)$. The $l_{1}$-magic toolbox \cite{Candes_Romberg_L1_Magic_Toolbox} provides BP and BPDN codes. 
In the absence of a slack estimate $\epsilon$, we can solve an $l_{1}$ norm penalized $l_{2}$ norm cost function in the following way:
\begin{eqnarray}
\def\argmin{\mathop{\rm{arg\hspace{2pt}min}}}
Q_{1}^{\lambda}: \hspace{0.5cm} \hat{\mathbf{x}} = \argmin_{\mathbf{x} \in \mathbb{R}^{N}} \hspace{3pt} \left\{ \lambda \| \mathbf{x} \|_{1} +  \| \mathbf{y} -  \mathbf{A} \mathbf{x} \|_{2}  \right\}.
\label{eq:CS_Standard_Problem_LASSO}
\end{eqnarray}
The above formulation is called LASSO where the regularization parameter $\lambda$ needs to be optimally chosen. The optimal choice of $\lambda$ can be performed by LARS (least angle regression) \cite{Efron_LARS}. SparseLab toolbox \cite{SparseLab_Toolbox} provides code for such an optimal LASSO. 

We note that BP/BPDN/LASSO does not provide a solution vector that is guaranteed to be $K$-sparse in any experimental condition. Therefore, to compute ASCE performance measure, we assume that the support set is constructed from the solution vector of BP/BPDN/LASSO by considering the atom-indices associated with $K$ highest amplitude coefficients. Then a new reconstructed signal $\hat{\mathbf{x}}$ is estimated using the orthogonal projection of $\mathbf{y}$ on the constructed support set of cardinality $K$ (LS solution). We used the new reconstructed signal for our purpose of evaluating SRNR. In the literature \cite{Haupt_Compressive_distilled_sensing}, \cite{Chatterjee_Sundman_Skoglund_2}, such an oracle driven approach is shown to provide better SRNR performance for convex relaxation algorithms. Through experimental evaluations, we also verified that the oracle driven approach improves the performance of BP/BPDN/LASSO significantly \cite{Chatterjee_Sundman_Skoglund_2}.

\subsection{Experimental Results} 
\label{subsec:Experimental_Results}

Using $N=500$, $K=20$, $S=100$ and $T=100$, we performed experiments. That means, we used 500-dimensional sparse signal vectors with sparsity level $K=20$. 
Such a $4\%$ sparsity level is chosen in accordance with real life scenarios, such as most of the energy of an image signal in the wavelet domain is concentrated within $2-4\%$ coefficients \cite{CS_introduction_Candes_Wakin}.
We used 100 realizations of $\mathbf{A}$ (i.e., $T=100$). For each realization of $\mathbf{A}$, we used 100 signal vectors that are randomly generated (i.e., $S=100$). Then, we incremented $\alpha$ from a lower limit to a higher limit in a small step-size (with the constraint that corresponding $M$ is an integer for a value of $\alpha$).
Therefore, for each CS method at a chosen $\alpha$, the performance is evaluated through averaging over $100 \times 100=10000$ realizations.

Let us first observe the effect of increasing $L$ for the POMP, LAOLS and SOLS algorithms. For the Gaussian sparse signal at clean measurement condition, we show the SRNR performance of the three algorithms in  Fig.~\ref{fig:Effect_of_L_on_AllOMPs_Gaussian_CleanMeasurement_1}. We show the results for $L=2$, 5 and 20, in the range of $\alpha$ from 0.1 to 0.2. In Fig.~\ref{fig:Effect_of_L_on_AllOMPs_Gaussian_CleanMeasurement_1} (c), we do not show the performance of SOLS for $L=2$ (for $L=2$, SOLS acts as POMP). From Fig.~\ref{fig:Effect_of_L_on_AllOMPs_Gaussian_CleanMeasurement_1}, we observe that the SRNR performance improves as $L$ increases and the performance improvement by increasing to $L=5$ from $L=2$ is similar to the performance improvement by increasing to $L=20$ from $L=5$. This observation corroborates that the performance improvement saturates with increasing $L$. 
Considering the user defined choice that $L \leq K$, we choose to use $L=20$ as it provides the best performance for all the three algorithms. 
For binary sparse signal, we also performed similar experiments and found the same trend in performance. 
In the later part of this subsection, we show the results for POMP, LAOLS and SOLS algorithms where $L=20$ is used. 

Next, we compared between OMP, OLS, SP, BP/BPDN, LASSO, POMP, LAOLS and SOLS algorithms. We first show the SRNR and ASCE performance results for Gaussian sparse signal in Fig.~(\ref{fig:Comparison_Using_Gaussian_Signal_3}). The results are shown for clean measurement condition and noisy measurement conditions with SMNR = 20 and 10 dB. Fig.~\ref{fig:Comparison_Using_Gaussian_Signal_3} (a) and (b) compare the performance of all the algorithms for clean measurement condition. The SRNR results of Fig.~\ref{fig:Comparison_Using_Gaussian_Signal_3} (a) is shown in Fig.~\ref{fig:Comparison_Using_Gaussian_Signal_3_SRNR_CleanMeasurement} for a better view. We observe that the OLS and OMP provide similar performance. BP provides better SRNR than LASSO, but similar ASCE. The SP performs poorer than the OLS and OMP algorithms. For the proposed algorithms, we find that all the three new algorithms perform better than the existing algorithms. POMP provides a considerable improvement over OMP, and even performs better than the BP. At $\alpha=0.2$, the POMP provides more than 6 dB SRNR performance improvement compared to OLS and OMP algorithms. The LAOLS outperforms all the other competing algorithms by a significant margin and can be considered the best. At $\alpha=0.2$, the LAOLS provides nearly 20 dB SRNR performance improvement over OLS and OMP algorithms and 15 dB improvement over BP. The SOLS and POMP are found to be the second and third best, respectively. Now, let us observe the performance trends for noisy measurement conditions. Fig.~\ref{fig:Comparison_Using_Gaussian_Signal_3} (c) and (d) show the comparative study at SMNR = 20 dB. In this case, we still find that the new algorithms are better than the existing algorithms. LAOLS and SOLS are found to show significant improvements than the others. A natural question is what happens if the noise power increases. At SMNR = 10 dB, Fig.~\ref{fig:Comparison_Using_Gaussian_Signal_3} (e) and (f) show the results where we note that the BPDN and LASSO provide similar performance trends and they are better than the other IGS algorithms. For the Gaussian sparse signal, the results show that the IGS algorithms are non-robust at a higher measurement noise power.

Now we show the comparative results for binary sparse signal in Fig.~(\ref{fig:Comparison_Using_Binary_Signal_3}). The results are shown for clean measurement condition and noisy measurement conditions with SMNR = 20 and 10 dB. Fig.~\ref{fig:Comparison_Using_Binary_Signal_3} (a) and (b) compare the performance of all the algorithms for clean measurement condition. We again observe that the OLS and OMP provide similar performance. However, in this case, the SP is found to perform better than the OLS and OMP algorithms. The convex relaxation algorithms (BP and LASSO) provide the best performance among all the algorithms. For the proposed algorithms, we find that POMP provides a considerable improvement over OMP. It can be said that LAOLS is the best performer among the IGS algorithms. Let us now compare the algorithms for noisy measurement conditions. At SMNR = 20 dB, Fig.~\ref{fig:Comparison_Using_Binary_Signal_3} (c) and (d) display the results where the trends are similar to the clean measurement case results shown in Fig.~\ref{fig:Comparison_Using_Binary_Signal_3} (a) and (b). However, the performance trends take an interesting turn at a higher noise power level. At SMNR = 10 dB, Fig.~\ref{fig:Comparison_Using_Gaussian_Signal_3} (e) and (f) show the results where we observe that the BPDN/LASSO does not provide the best performance, but the LAOLS provides the best performance closely followed by the second best performance of SOLS. For the binary sparse signal, the LAOLS, SOLS and SP are found to perform better than the BPDN/LASSO at a higher measurement noise power. 

Comparing all the algorithms for two different input signals at varying noise power and number of measurements, we note that the proposed POMP, LAOLS and SOLS have a promise to provide good performance. They consistently outperformed OMP and OLS algorithms. A specific example is that, for the Gaussian sparse signal at clean measurement condition, LAOLS provides 20 dB SRNR improvement than the OMP and OLS at $\alpha = 0.2$.

Next we performed a running time comparison between the competing algorithms for the CS reconstruction of Gaussian sparse signal at SMNR = 20 dB. The running time means the average execution time. To compute the running time, we used matlab codes executing on a desktop computer and the averaging is performed using CS execution times for 100 data (we used $T=10$ ans $S=10)$. We performed the running time experiments for different $(N,K)$ pairs. The chosen pairs are $(500,10)$, $(500,20)$, $(500,30)$, $(1000,20)$, $(1000,30)$ and $(1000,40)$. For each $(N,K)$ pair, the choice of $M = \lceil K \log N \rceil$. The running time results are shown in Table~\ref{table:RunningTimeComparison_of_POMP_LAOLS_SOLS}. Following the discussion in section~\ref{sec:Comparison_of_Computational_Complexity}, we denote the allowable $K$ by $K^{\star}$ for which the LAOLS and SOLS might have a lower complexity than the BPDN complexity (BPDN uses interior-point method). From Table~\ref{table:RunningTimeComparison_of_POMP_LAOLS_SOLS} we find that the OMP has the minimum running time. The POMP and SP have similar order of running time. The running time of LAOLS and SOLS increases heavily with the increase of $K$. In this table, SOLS has the parameter $\gamma=0.5$. Depending on the parameter $\gamma$, the SOLS running time can be adjusted. 
We performed the running time comparison for binary sparse signal also, but do not report the results as they are similar to the Gaussian sparse signal case.

\footnotesize
\renewcommand{\baselinestretch}{1}
\begin{table}[t]
\centering
\caption{Running time comparison}
\setlength{\tabcolsep}{3.5pt}

\begin{tabular}{c||ccc|ccc}
	\hline

	\multicolumn{1}{c||}{Algorithm}	&
	\multicolumn{3}{c|}{$N=500$}	& 
	\multicolumn{3}{c}{$N=1000$}	\\ \cline{2-7} 

	\multicolumn{1}{c||}{}	&
	\multicolumn{3}{c|}{$K^{\star}=15$}	& 
	\multicolumn{3}{c}{$K^{\star}=20$}	\\ \cline{2-7}
	  & $K=10$ & $K=20$ & $K=30$ & $K=20$ & $K=30$ & $K=40$  \\ \hline \hline
	\multicolumn{7}{c}{Existing algorithms}	\\ \hline 

        OMP & 0.0010 & 0.0023 & 0.0043 & 0.0042 & 0.0108 & 0.0220  \\
        OLS & 0.3076 & 1.0100 & 2.6124 & 1.0522 & 2.6973 & 6.0095  \\
        SP  & 0.0057 & 0.0193 & 0.0482 & 0.0226 & 0.0541 & 0.1063  \\
        BPDN& 0.4599 & 0.8140 & 1.1478 & 2.6550 & 2.9915 & 3.5564  \\
        LASSO&0.0442 & 0.1325 & 0.2700 & 0.2358 & 0.5379 & 1.1774  \\ \hline \hline
	
        \multicolumn{7}{c}{New algorithms}	\\ \hline

        POMP& 0.0098 & 0.0340 & 0.0945 & 0.0407 & 0.1099 & 0.2465  \\
        LAOLS&0.0752 & 0.6696 & 2.7874 & 1.1794 & 6.4881 & 22.1076 \\ 
        SOLS& 0.0337 & 0.2560 & 1.0172 & 0.4395 & 2.3020 & 7.6874  \\ \hline

	\end{tabular}
	
	\label{table:RunningTimeComparison_of_POMP_LAOLS_SOLS}
	
\end{table}
\renewcommand{\baselinestretch}{1}
\normalsize 



We end this section with interesting observations. We note that, depending on the signal statistics (Gaussian and binary), the algorithms show different performance trends. For example, LAOLS provides better performance for a Gaussian sparse signal in the clean and moderately noisy measurement conditions. On the other hand, BP/BPDN/LASSO provides better performance for a binary sparse signal in the same testing conditions. Another example is that the SP provides poor performance for the Gaussian sparse signal, but found to be good for the binary sparse signal. Therefore, a challenging research problem is to develop adaptive IGS algorithms which are robust to underlying signal statistics. 

\emph{Reproducible results:} In the spirit of reproducible results, we provide necessary download-able matlab codes in the following website: \linebreak https://sites.google.com/site/saikatchatt/softwares/. For CS reconstruction of Gaussian sparse signal at SMNR = 20 dB, the codes produce the SRNR and ASCE results shown in Fig.~\ref{fig:Comparison_Using_Gaussian_Signal_3} (c) and (d), and the running time comparison results shown in Table~\ref{table:RunningTimeComparison_of_POMP_LAOLS_SOLS}.

\begin{figure*}
\centering
\includegraphics[width=10in,height=6.2in,angle=90]{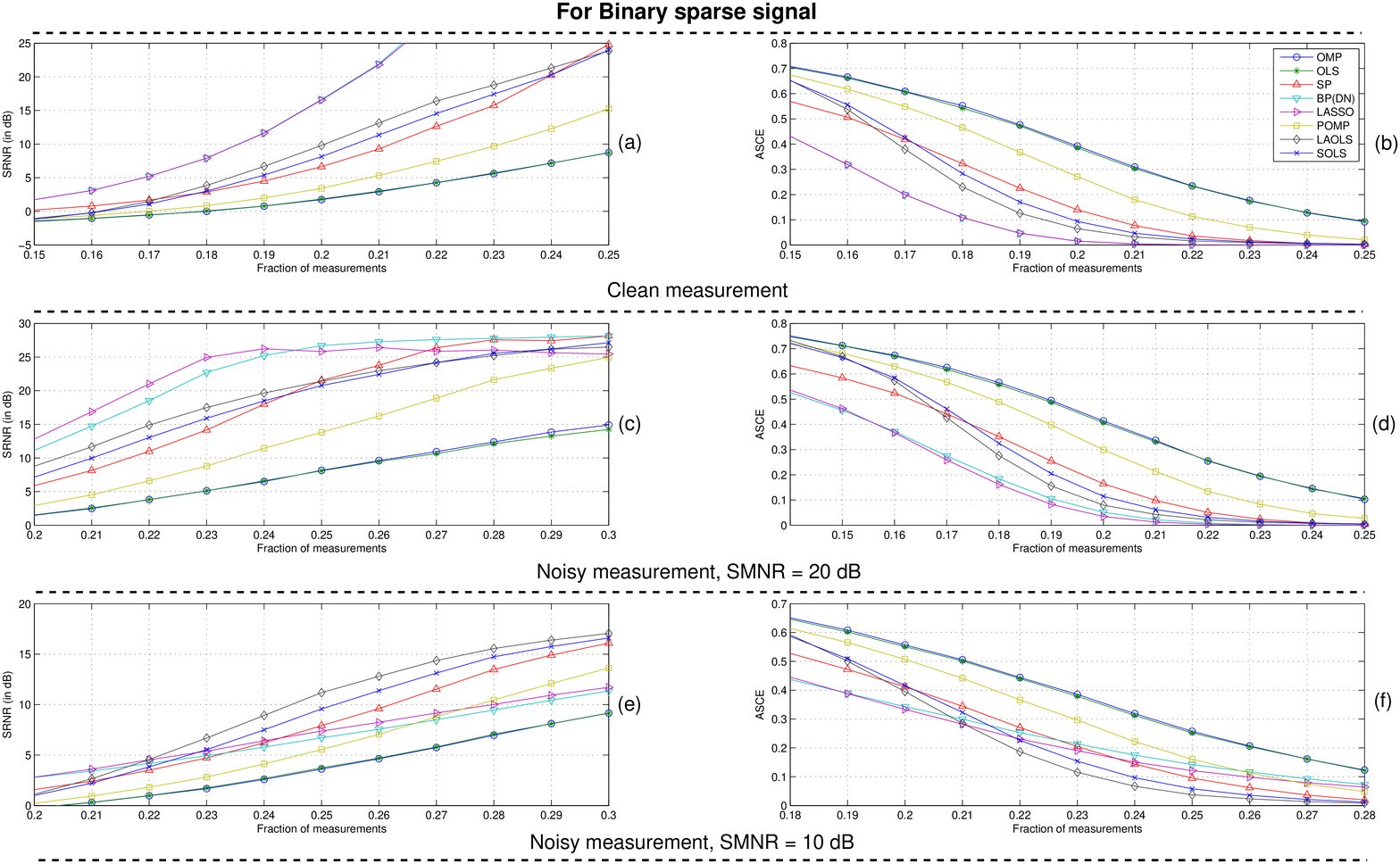}
\caption{Comparison of OMP, OLS, SP, BP/BPDN, LASSO, POMP, LAOLS and SOLS algorithms for {\bf{Binary sparse signal in clean and noisy measurement conditions}}. SRNR (in dB) and ASCE performances are plotted against the fraction of measurements ($\alpha$). We show the results for clean measurement as well as noisy measurements with SMNR = 20 and 10 dB. BP/BPDN is denoted by BP(DN).}
\label{fig:Comparison_Using_Binary_Signal_3}
\end{figure*}

\section{Discussions, Conclusions, and Future Works}

For greedy algorithms where atoms are selected one-by-one serially over iterations, we show that new schemes can be developed for selecting an atom from a set of potential atoms. The potential atoms are found using a standard matched filter. The use of standard matched filter reduces the search space reasonably and allows us to use more sophisticated tools for atom selection. Among the developed tools, the look ahead strategy allows us to see the evaluation of an algorithm in future iterations. Overall, for selecting atoms, the new tools help to improve further on the state-of-art performance of matched filter. 

Using our experimental framework, we also note that the performance of greedy search algorithms depend on the input signals' statistics. Future investigations can be performed to develop new greedy search algorithms that are robust to the statistics of input signals.

\section{Acknowledgement}
We acknowledge the authors of $l_{1}$-magic and SparseLab toolboxes for their generous distribution of codes online.

\appendix

\subsection{Serial Atom Selection Based OMP and OLS}
\label{subsec:Serial_Atom_Selection_Based_OMP_and_OLS}

In this section, we briefly describe two serial atom selection based IGS algorithms: OMP and OLS. First we summarize the main steps of the OMP in Algorithm~\ref{alg:OMP} (see \cite{Blumensath_Difference_on_OMP_and_OLS}, \cite{Blumensath_Gradient_pursuits} and Algorithm 3 of \cite{Tropp_OMP}).
\begin{algorithm}[ht!]
\caption{: OMP for CS Recovery}\label{alg:OMP}
\mbox{Input: }
\begin{algorithmic}[1]
\STATE $\mathbf{A}$, $\mathbf{y}$, $K$;
\end{algorithmic}
\mbox{Initialization: }
\begin{algorithmic}[1]
\STATE Iteration counter $k \leftarrow 0$;
\STATE $\mathbf{r}_{0} \leftarrow \mathbf{y}$, $\mathcal{I}_{0} \leftarrow \emptyset$;
\end{algorithmic}
\mbox{Iterations: }
\begin{algorithmic}[1]
\REPEAT 
\STATE $k \leftarrow k+1$;
\STATE $i_{k} \leftarrow $ index of the highest amplitude of $\mathbf{A}^{t} \mathbf{r}_{k-1}$; \label{step:atom_index_OMP}
\STATE $\mathcal{I}_{k} \leftarrow \mathcal{I}_{k-1} \cup i_{k}$; \hfill (Note: $|\mathcal{I}_{k}|=k$) \label{step:Intermediate_Support_OMP}
\STATE $\mathbf{r}_{k} \leftarrow \mathbf{y}-\mathbf{A}_{\mathcal{I}_{k}} \mathbf{A}_{\mathcal{I}_{k}}^{\dag} \mathbf{y}$; \hfill (Orthogonal projection)	\label{step:Proj_Residue_OMP}
\UNTIL $( (\| \mathbf{r}_{k} \|_{2} > \| \mathbf{r}_{k-1} \|_{2}) \,\, \mathrm{or}\,\, (k>K) )$
\STATE $k \leftarrow k-1$; \hfill (Previous iteration)
\end{algorithmic}
\mbox{Output:}
\begin{algorithmic}[1]
\STATE $\hat{\mathbf{x}} \in \mathbb{R}^{N}$, satisfying $\hat{\mathbf{x}}_{\mathcal{I}_{k}} = \mathbf{A}_{\mathcal{I}_{k}}^{\dag} \mathbf{y}$ and $\hat{\mathbf{x}}_{\overline{\mathcal{I}}_{k}} = \mathbf{0}$.
\end{algorithmic}
\end{algorithm}
Note that, in each iteration, the OMP performs a matched filter operation and an orthogonal projection. Using block-wise matrix inversion, the orthogonal projection operations can be performed recursively. Considering the worst case scenario that $k=K$, the necessary computation for each orthogonal projection is approximately $7K^2+4KM$. So, the total computation for each matched filter and each orthogonal projection is $\mathcal{O}(MN+K^2+KM)$. 
Now, considering $K$ iterations, the total complexity is $\mathcal{O}(K(MN+K^2+KM))$.

Next we summarize the OLS algorithm \cite{Blumensath_Difference_on_OMP_and_OLS} in Algorithm~\ref{alg:OLS}.
\begin{algorithm}[ht!]
\caption{: OLS for CS Recovery}\label{alg:OLS}
\mbox{Input: }
\begin{algorithmic}[1]
\STATE $\mathbf{A}$, $\mathbf{y}$, $K$;
\end{algorithmic}
\mbox{Initialization: }
\begin{algorithmic}[1]
\STATE Iteration counter $k \leftarrow 0$;
\STATE $\mathbf{r}_{0} \leftarrow \mathbf{y}$, $\mathcal{I}_{0} \leftarrow \emptyset$;
\end{algorithmic}
\mbox{Iterations: }
\begin{algorithmic}[1]
\REPEAT 
\STATE $k \leftarrow k+1$;
\STATE $\def\min{\mathop{\rm arg \,\, min}} i_{k} \leftarrow \min_{i: \mathcal{I}_{(u)} = \mathcal{I}_{k-1} \cup i} \| \mathbf{y} - \mathbf{A}_{\mathcal{I}_{(u)}} \mathbf{A}_{\mathcal{I}_{(u)}}^{\dag} \mathbf{y} \|_{2} $; \newline ($i \notin \mathcal{I}_{k-1}$ and orthogonal projections) \label{step:atom_index_OLS}
\STATE $\mathcal{I}_{k} \leftarrow  \mathcal{I}_{k-1} \cup i_{k}$; \hfill (Note: $|\mathcal{I}_{k}|=k$) \label{step:Intermediate_Support_OLS}
\STATE $\mathbf{r}_{k} \leftarrow \mathbf{y}-\mathbf{A}_{\mathcal{I}_{k}} \mathbf{A}_{\mathcal{I}_{k}}^{\dag} \mathbf{y}$; \hfill (Orthogonal projection)	\label{step:Proj_Residue_OLS}
\UNTIL $( (\| \mathbf{r}_{k} \|_{2} > \| \mathbf{r}_{k-1} \|_{2}) \,\, \mathrm{or}\,\, (k>K) )$
\STATE $k \leftarrow k-1$; \hfill (Previous iteration)
\end{algorithmic}
\mbox{Output:}
\begin{algorithmic}[1]
\STATE $\hat{\mathbf{x}} \in \mathbb{R}^{N}$, satisfying $\hat{\mathbf{x}}_{\mathcal{I}_{k}} = \mathbf{A}_{\mathcal{I}_{k}}^{\dag} \mathbf{y}$ and $\hat{\mathbf{x}}_{\overline{\mathcal{I}}_{k}} = \mathbf{0}$.
\end{algorithmic}
\end{algorithm}
The overall structure of the OLS algorithm is similar to the OMP algorithm. The only and important difference is the way how the atoms are chosen through iterations. In each iteration, an atom is selected in such a way that its choice leads to the minimum residual norm for that iteration. This approach of atom selection is different from the OMP which uses a matched filter to select an atom, but does not guarantee to provide the minimum residual norm. However, OLS is computationally more intensive than the OMP. The OLS requires to perform approximately $[N + (N-1) + \ldots + (N-K)]  = \frac{1}{2} (KN - K^2 + K)$ orthogonal projection operations. As typically $N \gg K$, the computational complexity can be significantly high. However, in this case also the orthogonal projection operations can be performed recursively using block-wise matrix inversion. Considering the worst case scenario that $k=K$, the necessary computation for each orthogonal projection is approximately $7K^2+4KM$. 
So, the overall complexity of OLS is $\mathcal{O}(KN(K^2+KM)) = \mathcal{O}(K^2(KN+MN))$.

\subsection{Parallel Atom Selection Based SP}
\label{subsec:Parallel_Atom_Selection_Based_SP}  
 
In this section, we describe the parallel atom selection based IGS algorithm: SP. We summarize the main steps of the SP algorithm in Algorithm~\ref{alg:SP} (see Algorithm 1 of \cite{Dai_Subspace_pursuit}).
\begin{algorithm}[ht!]
\caption{: SP for CS Recovery}\label{alg:SP}
\mbox{Input: }
\begin{algorithmic}[1]
\STATE $\mathbf{A}$, $\mathbf{y}$, $K$;
\end{algorithmic}
\mbox{Initialization: }
\begin{algorithmic}[1]
\STATE Iteration counter $k \leftarrow 0$;
\STATE $\mathcal{I}_{0} \leftarrow $ indices of the $K$ highest amplitudes of $\mathbf{A}^{t} \mathbf{y}$;
\STATE $\mathbf{r}_{0} \leftarrow \mathbf{y}-\mathbf{A}_{\mathcal{I}_{0}} \mathbf{A}_{\mathcal{I}_{0}}^{\dag} \mathbf{y}$;
\end{algorithmic}
\mbox{Iterations: }
\begin{algorithmic}[1]
\REPEAT 
\STATE $k \leftarrow k+1$;
\STATE $\mathcal{I}_{(p)} \leftarrow \left\{ \mathrm{indices \,\, of} \,\, K \,\, \mathrm{highest \,\, amplitudes \,\, of} \,\, \mathbf{A}^{t} \mathbf{r}_{k-1} \right\}$;
\STATE $\mathcal{I}_{(u)} \leftarrow \mathcal{I}_{k-1} \cup \mathcal{I}_{(p)}$; \hfill ($K \leq |\mathcal{I}_{(u)}| \leq 2K$)
\STATE $\hat{\mathbf{x}}_{\mathcal{I}_{(u)}} \leftarrow \mathbf{A}_{\mathcal{I}_{(u)}}^{\dag} \mathbf{y}$;
	$\hat{\mathbf{x}}_{\overline{\mathcal{I}}_{(u)}} \leftarrow \mathbf{0}$; \hfill (Orthogonal projection)
\STATE $\mathcal{I}_{k} \leftarrow $ $\{$indices of the $K$ highest amplitudes of $\hat{\mathbf{x}} \}$; 
\label{step:Intermediate_Support_SP} 
\STATE $\mathbf{r}_{k} \leftarrow \mathbf{y}-\mathbf{A}_{\mathcal{I}_{k}} \mathbf{A}_{\mathcal{I}_{k}}^{\dag} \mathbf{y}$; \hfill (Orthogonal projection)
\label{step:Proj_Residue_SP}
\UNTIL $( \| \mathbf{r}_{k} \|_{2} > \| \mathbf{r}_{k-1} \|_{2})$
\STATE $k \leftarrow k-1$; \hfill (Previous iteration)
\end{algorithmic}
\mbox{Output:}
\begin{algorithmic}[1]
\STATE $\hat{\mathbf{x}} \in \mathbb{R}^{N}$, satisfying $\hat{\mathbf{x}}_{\mathcal{I}_{k}} = \mathbf{A}_{\mathcal{I}_{k}}^{\dag} \mathbf{y}$ and $\hat{\mathbf{x}}_{\overline{\mathcal{I}}_{k}} = \mathbf{0}$.
\end{algorithmic}
\end{algorithm}
The SP algorithm starts with an initial $K$-element support set $\mathcal{I}_{0}$ and an initial residual $\mathbf{r}_{0}=\mathbf{y}-\mathbf{A}_{\mathcal{I}_{0}} \mathbf{A}_{\mathcal{I}_{0}}^{\dag} \mathbf{y}$. At the $k$'th iteration stage, it forms the `matched filter' $\mathbf{A}^{t}\mathbf{r}_{k-1}$, identifies the $K$ highest amplitude coordinates, forms a dummy support set $\mathcal{I}_{(u)} = \mathcal{I}_{k-1} \cup \mathcal{I}_{(p)}$, refines out $K$-element support set $\mathcal{I}_{k}$ from $\mathcal{I}_{(u)}$, solves a LS problem with the selected indices in $\mathcal{I}_{k}$, subtracts the LS fit and produces a new residual. Given the sparsity level $K$, the algorithm estimates a support set of cardinality $K$ in each iteration and runs until the residual norm minimization condition is violated. Note that, unlike in the case of serial atom selection based OMP and OLS algorithms, here the support set cardinality is not increased one-by-one through iterations. Rather, a $K$-element support set is refined through iterations by addition of potential new atoms and deletion of unnecessary atoms. An important point is to note how the $K$-element support set $\mathcal{I}_{k}$ is chosen from the dummy support set $\mathcal{I}_{(u)}$ through using the orthogonal projection that invokes LS solution. The dummy support set $\mathcal{I}_{(u)}$ is formed through unionizing the previously estimated support set $\mathcal{I}_{k-1}$ with the set of $K$ new atoms' indices. Then the observation $\mathbf{y}$ is orthogonally projected on the span of atoms that are indexed in $\mathcal{I}_{(u)}$ followed by picking up $K$ indices corresponding to the highest amplitude coefficients of the solution vector.  
In the context of POMP/SOLS algorithm, we used similar strategy in Algorithm~\ref{alg:Proj_atom_selection}/Algorithm~\ref{alg:Proj_Multiple_atom_selection} for improving the atom selection strategy where a single atom/multiple atoms is/are selected from a set of potential atoms in each iteration. 
Normally, the SP algorithm converges less than $K$ iterations. Assuming the worst case scenario that the SP algorithm runs at most $K$ iterations, it performs $K$ matched filtering and $2K$ orthogonal projection operations. Note that, like OMP and OLS, the orthogonal projections can not be performed recursively. Hence, the total complexity for each iteration is $\mathcal{O}(MN + K^2M)$. Therefore, considering $K$ iterations, the overall complexity of SP algorithm is $\mathcal{O}(K(MN + K^2M))$.

\subsection{Recursive Computation for Look Ahead Strategy}
\label{subsec:Resursive_Computation_for_Look_Ahead}
In Algorithm~\ref{alg:LA_strategy}, a computationally intensive part is performing the orthogonal projection operation in each iteration. For orthogonal projection, we need to perform pseudo-inverse of $\mathbf{A}_{\mathcal{I}_{k}} \in \mathbb{R}^{M \times k}$ where $\mathbf{A}_{\mathcal{I}_{k}} = \left[ \mathbf{A}_{\mathcal{I}_{k-1}} \,\, \mathbf{a}_{i_{k}} \right]$. The pseudo-inverse $\mathbf{A}_{\mathcal{I}_{k}}^{\dag} \mathbf{y} = \left[ \mathbf{A}_{\mathcal{I}_{k}}^{t} \mathbf{A}_{\mathcal{I}_{k}} \right]^{-1} \mathbf{A}_{\mathcal{I}_{k}}^{t} \mathbf{y}$ and we note that the computation of inverse operation can be performed recursively. Denoting $\mathbf{P}_{k}^{-1} = \left[ \mathbf{A}_{\mathcal{I}_{k}}^{t} \mathbf{A}_{\mathcal{I}_{k}} \right]^{-1}$, we show (using block-wise matrix inversion) the recursive expression in~(\ref{eq:Recursive_Matrix_1})
\begin{figure*}
\hrule
\begin{eqnarray}
\begin{array}{rcl}
\left[ \mathbf{P}_{k} \right]^{-1}  & = & \left[   
\left[
\begin{array}{cc}
\mathbf{A}_{\mathcal{I}_{k-1}}^{t} \mathbf{A}_{\mathcal{I}_{k-1}}  & \mathbf{A}_{\mathcal{I}_{k-1}}^{t} \mathbf{a}_{i_{k}} \\
\mathbf{a}_{i_{k}}^{t}\mathbf{A}_{\mathcal{I}_{k-1}}     & \mathbf{a}_{i_{k}}^{t}\mathbf{a}_{i_{k}}
\end{array}
\right] \right]^{-1}  = \left[
\begin{array}{cc}
\mathbf{P}_{k-1} & \mathbf{q}_{k}  \\
\mathbf{q}_{k}^{t}  & 1  
\end{array}
\right]^{-1}   \\
 & = & \left[
\begin{array}{cc}
\left[ \mathbf{P}_{k-1} \right]^{-1}  + \frac{1}{\beta_{k}} \! \left[ \mathbf{P}_{k-1} \right]^{-1} \mathbf{q}_{k}  \left[ \left[ \mathbf{P}_{k-1} \right]^{-1} \mathbf{q}_{k} \right]^{t} & - \frac{1}{\beta_{k}} \left[ \mathbf{P}_{k-1} \right]^{-1} \mathbf{q}_{k}    \\
- \frac{1}{\beta_{k}}  \left[ \left[ \mathbf{P}_{k-1} \right]^{-1} \mathbf{q}_{k} \right]^{t}  & \frac{1}{\beta_{k}} 
\end{array}
\right]
\end{array}
\label{eq:Recursive_Matrix_1}
\end{eqnarray}
\hrule
\end{figure*}
where $\mathbf{q}_{k}=\mathbf{A}_{\mathcal{I}_{k-1}}^{t} \mathbf{a}_{i_{k}}$ (note $\mathbf{q}_{k} \in \mathbb{R}^{k-1}$) and the scalar $\beta_{k}= 1 - \mathbf{q}_{k}^{t} \left[ \mathbf{P}_{k-1} \right]^{-1} \mathbf{q}_{k}$. Here we use the fact that $\mathbf{P}_{k}$ is symmetric. Also note that the computation of $\mathbf{A}_{\mathcal{I}_{k}}^{t} \mathbf{y}$ can be performed recursively as 
\begin{eqnarray}
\mathbf{A}_{\mathcal{I}_{k}}^{t} \mathbf{y} = \left[
\begin{array}{c}
\mathbf{A}_{\mathcal{I}_{k-1}}^{t} \mathbf{y} \\
\mathbf{a}_{i_{k}}^{t}\mathbf{y}
\end{array}
\right].
\end{eqnarray}  
Now, considering the fact that the multiplication of two matrices, $\mathbf{C} \in \mathbb{R}^{m \times n}$ and $\mathbf{D} \in \mathbb{R}^{n \times k}$, requires $2kmn$ operations, we can show that the total required computation for pseudo-inverse of $\mathbf{A}_{\mathcal{I}_{k}} \in \mathbb{R}^{M \times k}$ is $\approx 7k^2+4kM$.

%








\end{document}